\documentstyle[11pt,aaspp]{article}

\def\asca{{\it ASCA}}
\def\axaf{{\it AXAF}}
\def\einstein{{\it Einstein}}
\def\EO{{\it Einstein Observatory}}
\def\rosat{{\it ROSAT}}
\def\hst{{\it HST}}
\def\hii{\ion{H}{2}}
\def\kms{km s$^{-1}$}

\def\myarcmin{^\prime\mskip-5mu}

\def\lsim{\hbox{\raise.35ex\rlap{$<$}\lower.6ex\hbox{$\sim$}\ }}
\def\gsim{\hbox{\raise.35ex\rlap{$>$}\lower.6ex\hbox{$\sim$}\ }}

\begin{document}

\title{{\it ASCA} X-Ray Spectroscopy of LMC Supernova Remnants and \\
the Metal Abundances of the LMC}

\author{John P. Hughes}
\affil{Department of Physics and Astronomy, Rutgers
University, \\ 136 Frelinghuysen Road, Piscataway, NJ 08854-8019 \\ 
jph@physics.rutgers.edu}

\vskip 10 pt

\author{Ichizo Hayashi and Katsuji Koyama}
\affil{Department of Physics, Graduate School of Science, Kyoto University, \\
Sakyo-ku, Kyoto 606-8502, Japan \\ 
hayashi@cr.scphys.kyoto-u.ac.jp, koyama@cr.scphys.kyoto-u.ac.jp} 

\begin{abstract}

We present the results of X-ray spectroscopy of a flux-limited sample
of seven middle-aged supernova remnants (SNRs) in the Large Magellanic
Cloud (LMC): N23, N49, N63A, DEM71, N132D, 0453$-$68.5, and N49B.  We
constructed self-consistent nonequilibrium ionization SNR models
assuming a Sedov solution for the dynamical evolution,
and then applied the resulting spectral models to the data obtained by
the Solid-state Imaging Spectrometer onboard the {\it Advanced
Satellite for Cosmology and Astrophysics}. All the remnants were
reasonably well described by the model, allowing us to derive accurate
values for their physical parameters, i.e., ages, densities, initial
explosion energies, and metal abundances.  The derived explosion
energies vary from $5\times 10^{50}$ erg to $6\times 10^{51}$ erg. A
restricted subset of the sample exists for which the ionization and
Sedov dynamical ages agree quite well under the assumption that the
electron and ion temperatures are not fully equilibrated at the shock
front; for these four SNRs the mean value of the initial explosion
energy is $(1.1\pm 0.5)\times 10^{51}$ erg. We show that it is likely
that the other three remnants exploded within preexisting cavities in
the interstellar medium.  
The limits on high energy X-ray emission 
($\gsim$3 keV) that we present indicate that these SNRs do not contain
very luminous pulsar-powered synchrotron nebulae, in general agreement
with our picture of them as evolved, middle-aged remnants.
We find statistical evidence for enrichment
by supernova ejecta in the sense that smaller remnants show a somewhat
higher mean metallicity than the larger ones.  In the case of DEM71,
the putative remnant of a Type Ia supernova, the derived abundance of
iron is about a factor of two larger than the other remnants in the
sample. These things being said, however, the derived abundances are
in general dominated by swept-up interstellar material and so we use
the SNR sample to estimate the mean LMC gas-phase abundances.  We find
that the astrophysically common elements from oxygen to iron are less
abundant than the solar values by factors of 2--4. Overall these
results are consistent with previous ones based on optical and UV
data, but do not show the anomalous overabundance of magnesium and
silicon seen by others.

\end{abstract}


\section{Introduction}
  
Studies of the supernova remnants (SNRs) in the Large Magellanic Cloud
(LMC) are essential for elucidating the details of SNR evolution,
nucleosynthesis, the nature and environments of supernova progenitors,
and so on. As probes of the interstellar medium (ISM), SNRs also yield
information on the energy balance and chemical composition of their
environments, which serves as input to larger questions of galactic
chemical evolution and the star formation history of the Cloud.  The
relative closeness of the LMC plus its well-determined distance (we
use 50 kpc throughout) means that accurate physical quantities can be
derived for individual remnants in the galaxy. Also, due to the
generally low interstellar absorption toward the LMC, it is possible
to detect X-rays in the important 0.5--2.5 keV energy band, which
includes emission lines from some of the most abundant metals in the Universe,
such as K-shell transitions of highly-ionized atoms of oxygen, neon,
magnesium, silicon and sulfur and L-shell lines of ionized iron.  The
Solid-state Imaging Spectrometer (SIS) onboard the {\it Advanced
Satellite for Cosmology and Astrophysics} (\asca), with its superior
energy resolution compared to previous broadband X-ray spectrometers,
gives us the first real opportunity for studying these issues
quantitatively in the X-ray regime.

Mathewson et al.\ (1983, 1984, 1985) identified 32 SNRs in the LMC on
the basis of optical, radio, and X-ray observations.  The remnants
range in size from 2 pc to about 100 pc, and therefore span the range
of evolutionary phases, from young ejecta-dominated remnants, through
middle-aged adiabatic remnants, to old remnants in the radiative
phase.  At least 12 of the LMC remnants are bright enough X-ray
emitters ($F_{X} > 10^{-11}$ erg s$^{-1}$ cm$^{-2}$ in the
0.15--4.5 keV band), that we can perform detailed spectroscopic study
with \asca.  First results have already been reported from \asca\ on
three small LMC remnants, N103B, 0509$-$67.5, and 0519$-$69.0 (Hughes et
al.\ 1995), that show strong emission lines originating from
ejecta characteristic of thermonuclear supernovae, i.e., Type Ia SNe.

For the research described herein, we observed seven middle-aged
remnants, N23, N49, N63A, DEM71, N132D, 0453$-$68.5, and N49B, with
\asca\ and performed detailed spectral analysis of their SIS data.
This is all the LMC SNRs larger than 5 pc (radius) and brighter than
$10^{11}$ ergs s$^{-1}$ cm$^{-2}$ over the soft keV X-ray band,
which includes the three brightest thermal X-ray--emitting remnants in
the LMC (N132D, N63A, and N49), plus four less luminous ones that
cover a range of evolutionary states.  Previous X-ray spectroscopic
studies of N49, N63A, N132D, and N49B were reported by Clark et al.\
(1982) based on Solid State Spectrometer (SSS) data from the \EO. The
SSS data clearly showed weak emission lines of Mg, Si, and S
indicative of thermal emission, although the analysis, which used
somewhat unphysical two-temperature collisional ionization equilibrium
plasma models, left some doubt about the reliability of the derived
elemental abundance values.  More recently Hwang et al.\ (1993)
analyzed all the available \einstein\ data on N132D using a
phenomenological nonequilibrium ionization (NEI) model and derived
abundances of the heavy elements that were considerably lower than the
solar values. Both analyses suggest that the bulk of the X-ray
emission from these remnants comes from swept-up ISM rather than
supernova ejecta and, as we show below, our \asca\ data confirm this
finding.  This observational fact gives us confidence that the SNR
abundances we measure indeed correspond to those of the ISM in the
LMC.

The elemental abundances of the Magellanic Clouds are of great
importance to extragalactic astronomy.  As the nearest major galaxies
to our own, they are well-studied and serve as the stepping stones for
studies of more distant systems. In the current chemical composition
of the Clouds we see at least some of the material returned to the
ISM by stars of all types through winds and supernovae and, using
models of stellar evolution and nucleosynthesis, we can constrain the
history of star formation in these galaxies.  It is well known that
the Clouds have undergone quite different star formation histories
from the Galaxy, based on their differing chemical properties (see,
for example, Tsujimoto et al.~1995).

In the past 20 years with the advent of high quality optical and UV
spectra, the study of the chemical abundances of the LMC has made
great progress.  The earlier results of this era (summarized in Dufour
1984) showed that the metallicity of LMC \hii\ regions was
significantly less than solar; the observed oxygen abundance, for
example, was only about \onethird\ the solar value.  More recent
studies (Luck \& Lambert 1992, Russell \& Dopita 1992, hereafter RD92)
have, in general, tended to confirm these findings using optical
spectra and detailed modeling of \hii\ regions, SNRs, Cepheids, and
supergiant stars in the LMC.  However, even though considerable effort
has been expended in this area of research over the past decade, there
remains considerable room for improvement.  In particular, certain
critical elements, such as Mg and Si, are not easily studied in the
optical and UV bands due to a paucity of appropriate atomic
transitions. The work we present here represents the first systematic
attempt to measure the gas-phase LMC abundances in a new, independent
manner using X-ray spectroscopy of SNRs. 

Our spectral analysis also allows us, for the first time, to derive
accurate values for the temperature, ionization timescale, and
intrinsic X-ray emissivity of the hot gas in LMC SNRs.  When combined
with their physical radii, which are accurately known because of the
well-established distance to the LMC, we have four independent
observational constraints on the dynamical evolution.  Only three of
these quantities are necessary to fully specify the evolutionary state
according to the Sedov (1959) similarity solution, so in fact the
observational data result in an overconstrained system.  This allows
us to go beyond merely assuming the Sedov solution and instead allows
to actually establish whether or not the Sedov model is a correct
representation for the evolution of the SNRs in our sample.

In the following we present our \asca\ analysis of the selected LMC
SNR sample. In \S 2 we briefly present the observational details.  In
contrast to the phenomenological models used in previous X-ray studies
of LMC SNRs, we have developed realistic self-consistent models that
incorporate the dynamical evolution of remnants.  Section 3 provides a
description of the spectral model we developed and compares our
calculations to similar published ones. Results of the Sedov model
fits, as well as considerations of the applicability of the model, are
given in \S 4.  Discussions of the derived abundances are in
\S 5.  In \S 6 we examine the high energy X-ray emission ($\gsim$3
keV) of the remnant sample. We comment on individual remnants in
\S 7 and summarize in the final section.

\section{Observations}

\asca\ has four identical X-ray telescopes (XRT) and four focal plane
detectors.  Two of the focal plane instruments are solid-state imaging
spectrometers (SIS0 and SIS1; CCD cameras) and the other two are gas imaging
spectrometers (GIS2 and GIS3; gas scintillation proportional counters).  A
summary of the \asca\ observatory and its complement of instruments has been
given by Tanaka, Inoue, \& Holt (1994).

The log of observations for our remnant sample is given in Table 1.
All but one of the remnants were observed during the Performance
Verification phase or the first Guest Observer cycle. The seven remnants
in our sample were all found to have soft spectra (mean electron
temperatures $kT_{e}\lesssim 1$ keV), so we used only the SIS data for
the present spectral analysis.  The SIS was operated in 1-CCD faint
mode for all observations.  The data were corrected for dark-frame
error, the echo effect, and charge-transfer inefficiency and hot and
flickering pixels were removed using standard procedures.  Data were
rejected during orbital times of high background (South Atlantic
Anomaly passages) or when the geomagnetic cutoff rigidity was less
than 6 GeV\,c$^{-1}$. In order to avoid contamination due to light
leaks through the optical blocking filters, we excluded all data taken
when the satellite viewing direction was 10$^{\circ}$ or less from the
bright rim of the earth.  X-ray spectra were extracted from circular
regions of radius $4^\prime$ centered on each SNR with the exception
of N49 and N49B.  Due to their close relative proximity (N49 and N49B
are separated on the sky by merely $6\farcm7$), the spectra of these
two remnants were obtained from $3^\prime$ radius regions.  In all
cases background spectra were constructed from source-free regions in
the SIS field of view of the same observation. We show the
background-subtracted SIS0 spectra of our remnant sample in Figure \ref{fig1}.

\section{Spectral Models}

\subsection{Phenomenological Model}
\label{phenomenological}

The X-ray spectra in Figure \ref{fig1} are all clearly thermal in nature,
showing characteristic K-shell line emission from highly ionized atoms
of oxygen, neon, magnesium, silicon, and sulfur as well as a broad
blend of iron L-shell emission near 1 keV.  The spectra of the
different SNRs are remarkably similar to each other, while they differ
sharply from the young ejecta-dominated remnants which we studied in
our earlier work on the LMC SNRs (Hughes et al.~1995). In fact,
whereas the \asca\ spectra of the ejecta-dominated SNRs appear
qualitatively more consistent with an abundance distribution
corresponding to Type Ia SN ejecta, the seven remnants in the current
study are more consistent with the roughly \onethird\ solar abundances
of the LMC.

As our first approach to a quantitative analysis of these spectra we
used a phenomenological model, free from any specific dynamical model
of SNR evolution, for including NEI effects in the spectral fitting
(Hayashi et al.~1995).  In this approach the X-ray-emitting plasma is
assumed to have been shock heated some time ago $t$ to a constant
temperature $T_{e}$.  The ionization timescale, which is the parameter
characterizing deviations from ionization equilibrium, is given by
$n_{e}t$, where $n_{e}$ is the mean electron density. In simple terms
these parameters are determined from the data in the following manner:
$T_{e}$ is largely constrained by the shape of the continuum emission
while $n_{e}t$ is constrained by the emission line ratios (principally
the ratio of helium-like to hydrogen-like K$\alpha$ lines). In addition to
the temperature and ionization timescale, the elemental abundances, an
emission measure, and the line-of-sight absorption column density were
required to completely specify the model.  The code is based on the
NEI model from Hughes \& Helfand (1985); we use the implementation
described in Hughes \& Singh (1994).

No remnant could be adequately described by a single-component NEI
plasma model, i.e., with one $T_{e}$ and $n_{e}t$ for all species.  In
particular, the degree of ionization appeared to vary from element to
element.  Similar discrepancies of this kind were found in fits of the
phenomenological NEI plasma model to the \asca\ data of the young
oxygen-rich SNR 0102$-$72.3 in the Small Magellanic Cloud (Hayashi et
al.~1994). We therefore modified our fits to the LMC SNRs to allow for
variable ionization timescales for the different elemental species
displaying prominent line emission in the observed band, while at the
same time retaining a single value of $T_{e}$. This model provided
improved fits, the results of which allowed us to estimate the
quantity of X-ray--emitting plasma in the various SNRs.  Assuming a
constant density shell filling \onequarter\ of the entire volume,
using the fitted elemental abundances (which were all substantially
less than solar, Hayashi et al.~1995), and taking the remnant radii
from Mathewson et al.~(1983) (see also Table 2), we obtained masses
ranging from several tens to several hundred solar masses.  Clearly
these are evolved remnants that have swept up a large quantity of ISM.

Nevertheless, the apparent dependence of $n_{e}t$ on elemental species as
implied by this phenomenological NEI model, makes us question its physical
plausibility.  Taken at face value, one interpretation of the ionization
timescale variation might be that the various elemental species are in
spatially disjoint zones that have undergone differing thermodynamic
histories.  Since the amount and approximate composition of the
X-ray--emitting plasma are consistent with being dominated by ISM material
(which should be chemically homogeneous), rather than SN ejecta (which may not
be), we are confident that this explanation can be rejected.  An alternate and
indeed more physically plausible explanation is that the X-ray--emitting
plasma is chemically homogeneous, but contains a broad range of temperatures
and ionization timescales. This situation would arise as a natural consequence
of incorporating realistic models for the dynamical evolution of SNRs, since
our spectra are integrated over the entire X-ray emission from each
remnant. Apparent variation of the average value of $n_{e}t$ with composition,
for example, would then arise as a result of the strong dependence of
ionization, recombination, and excitation rates, as well as other atomic
physics quantities, on elemental species.  In effect, each species would tend
to pick out that value of $n_{e}t$ which produces a maximum of intensity in
the relevant emission lines.  Some support for this scenario comes from the
fact that, at least for N132D, a better fit to the \asca\ data was obtained
using a two component phenomenological NEI model that contained emission from
gas at both $\sim$0.5 keV and $\sim$2 keV (Hayashi 1997) (see \S\ref{n132d}
below).

\subsection{Sedov Model}
\label{sedovmodel}

We consider the Sedov (1959) similarity solution for an adiabatic blast wave
expanding into a uniform, homogeneous ISM as the description for the dynamical
evolution of our SNRs.  As is well known this phase of evolution begins when
the mass of swept-up ISM grows to be much more than the mass of SN ejecta (so
that the ejecta are dynamically insignificant) (Gull 1973) and it lasts until
radiative cooling becomes important, which occurs roughly when the shock
temperature drops below $\sim$$10^6$ K (Cox 1972). Based on our estimates of
the swept-up mass from the phenomenological model fits, the seven LMC SNRs in
the sample are all old enough that they satisfy the first constraint and thus
are likely to be in the adiabatic phase of evolution.  In \S\ref{results}
below we show that the remnants are not so old that radiative cooling has
affected their dynamical evolution, and so they also satisfy the second
condition.

In order to calculate the emergent NEI X-ray spectrum, the time
history of the electron density and temperature throughout the model
remnant is required.  For this calculation we divided the remnant into
a number of spherically symmetric shells spaced linearly in the
nondimensional variable $V = u/(r/t)$, where $u$ and $r$ are the
velocity and radial position of a specific shell, respectively, and
$t$ is the remnant's age.  This choice of integration variable ensures
a fine spacing near the shock front where the emission measure of the
remnant is the largest and the time variation of the relevant
quantities is the most rapid. For simplicity, the time integration
was constrained so that on successive time steps a single new spatial
shell was shock heated and incorporated into the remnant. Consequently
the more finely the remnant was divided spatially, the shorter was the
time step. Based on some trial studies with the \asca\ spectrum of
N132D, it was found that 30 shells provided an optimal compromise
between execution speed and calculational accuracy and so this value
was adopted for the analysis. The temperature and density of each
interior shell at each time step was determined from the analytical
expression for the interior variation of these quantities (Sedov
1959).  It was assumed that the composition was spatially uniform and
that hydrogen and helium were fully ionized.  The mean molecular
weight per particle was also assumed constant across the remnant,
which is a good approximation for the low metallicity environment of
the LMC, where the gas is dominated by hydrogen and helium.

With the time dependence of the electron density and temperature now
fully specified, the ionization rate equations were integrated using
the matrix diagonalization method (e.g., Masai 1984; Hughes \& Helfand
1985) to determine the final (observed) ionization state as a function
of position.  Then the radial run of temperature, coupled with the
ionization state, was used to calculate the total emergent X-ray
spectrum summed over the entire model SNR.  We used the ionization and
recombination rate coefficients and the X-ray emission code of Masai
(1984).  The unshocked ISM was assumed to be in ionization equilibrium
at a temperature of $T=3\times 10^5$ K, but the final results are
very insensistive to the precise value of this quantity. Again
experiments with the spectrum of N132D found no significant difference
in the best-fit parameters when the temperature defining the initial
equilibrium ionization fractions was varied from $10^4$ K to $10^6$ K.
This is because ionization of the first few ion states proceeds very
rapidly; most of the time is spent in the helium- and hydrogen-like
ionic stages for the medium $Z$ atomic species that are important in
our analysis (i.e., the abundant elements from oxygen to sulfur, see
below).

Emission from thirteen elements (H, He, C, N, O, Ne, Mg, Si, S, Ar,
Ca, Fe, and Ni) is included in our model.  The abundances of helium
and carbon cannot be determined from our data since there are no
significant emission lines from these species in the \asca\ X-ray
band.  Furthermore, we found that the abundances of nitrogen, argon,
calcium, and nickel were not well constrained by the data.
Consequently we fixed the abundances of these elements to the LMC
values determined by RD92: He 10.94, C 8.04, N 7.14, Ar 6.29, Ca 5.89,
and Ni 6.04, where we quote values on a scale of $12 + \log ({\rm
M/H})$. The baseline abundances we assume for the other species are O
8.82, Ne 7.92, Mg 7.42, Si 7.52, S 7.2, and Fe 7.6, which are the
cosmic values from Allen (1973).  The precise values of the carbon,
nitrogen, argon, calcium, and nickel abundances have virtually no
effect on our measurements of the other elemental abundances, unless
the listed species are extremely overabundant relative to solar, which
is unlikely, again given the low metallicity of the LMC.  The helium
abundance does have some impact on the absolute values of our derived
abundances, since a significant fraction of the continuum emission
(roughly 30\% at 1 keV) is provided by helium.  Roughly speaking a
10\% increase (decrease) in the helium abundance translates into a
5-10\% increase (decrease) in the derived abundances of the other
elements.  On the other hand, the relative abundance values we derive
are affected to a much lesser extent by changes in the helium
abundance.

The parameters of the model are the shock temperature ($T_{s}$), the
ionization timescale ($n_{0}t_{i}$, where $n_{0}$ is the preshock ISM
number density of hydrogen atoms and $t_{i}$ is the ionization age of
the remnant), the abundances of the elements listed above, and a
normalization factor ($N = n_{0}^{2}\theta_{R}^{3}D$, where
$\theta_{R}$ and $D$ are the angular radius and distance of the
remnant, respectively). The quantities $T_{s}$ and $n_{0}t_{i}$ are
constrained by the data in the same manner as described above for the
phenomenological NEI model and the normalization factor is essentially
determined from the X-ray flux.  When the radius of each remnant is
included, we have four independent observational constraints on the
three independent parameters of the Sedov model, namely age, ambient
ISM density, and initial explosion energy.  This overconstrained
system allows us to carry out an internal check on whether or not the
Sedov solution is a correct description of the SNRs in our sample.  In
effect what we do is to explore the consistency between the age
derived from the dynamics of the Sedov model (which depends on $T_{s}$
and $\theta_{R}$) and that derived from the fitted ionization
timescale (converted to an age using $n_{0}$ derived from the
normalization factor). One complication is that none of the remnants
in our sample displays the ideal, spherical symmetry expected of the
Sedov model, so an additional multiplicative factor is needed to
account for the fraction of volume within the radius of the remnant
that is filled with X-ray emitting gas when the fitted normalization
factor is converted to a mean density.  The last parameter in the
spectral fits is the line-of-sight absorbing column density $N_{\rm
H}$.  We use the cross sections of Morrison \& McCammon (1983).

The precise relationship between the amount of energy imparted to ions
and electrons at supernova shock fronts is still one of the major
unsolved issues in the physics of strong collisionless shocks
(Shklovsky 1968). Whether the electrons attain the same mean thermal
velocity behind the shock front as the ions, in which case they will
have a considerably lower temperature, or very rapidly equilibrate
with the ions, as proposed by McKee (1974), is still not entirely
clear. Therefore we have constructed two classes of models that
encompass the plausible extremes of these possible situations.  The
first is a full equilibration model, where the electron and ion
temperatures are assumed to be equal. The other model assumes that the
electrons and ions attain the same mean thermal velocity at the shock
and then share energy solely through Coulomb collisions (see Spitzer
1978) and that, consequently, the electron temperature lags the ion
temperature (see, for example, Itoh 1978; Cox \& Anderson
1982). However, as it turns out for our sample of LMC remnants, with
the possible exception of the ionization timescale, the fitted
parameters do not depend strongly on which model is assumed and, in
particular, the derived abundances are almost the same in the two
models.

As part of our model verification we computed a number of simple diagnostic
quantities and compared our results with those from the similar Sedov-phase
NEI models of Hamilton, Sarazin, \& Chevalier (1983).  We found reasonable
agreement (within a factor $\simeq2$) between the two models for the continuum
flux at 2 keV, the He-like K$\alpha$ line flux, the Fe-L line flux, and the
X-ray luminosity above 0.1 keV. Our model gives a slightly higher ionization
state than the Hamilton et al.\ model, which is most likely due to the
difference in the rate coefficients used by the models.

\section{Results of Sedov Model Fits}
\label{results}

In Figure \ref{fig1} we show the SIS spectra and the best-fit NEI 
spectral models for all the remnants in our sample. Sedov models assuming 
full electron-ion temperature equilibration are plotted; the other 
set of models assuming Coulomb temperature equilibration would be 
indistinguishable from these.  For display purposes only we averaged 
the SIS0 and SIS1 data and corresponding best-fit spectral models.  
For the actual analysis, in all cases but one, both the SIS0 and
SIS1 spectra were fitted jointly.  The one exception is N63A,
where there was a pronounced gain discrepancy for the SIS1 data.
Rather than introducing an {\it ad hoc} gain correction we decided to
use only the SIS0 data for this SNR.  Because of its high flux and
reasonably long exposure (see Table 1), N63A has the best statistics
of any remnant in the sample and so the loss of the SIS1 data does not
seriously impede our ability to determine its spectral parameters.  In
all cases the SIS data were rebinned so that each spectral channel
contained at least 20 counts after background subtraction.  This
allows us to use $\chi^2$ as the figure-of-merit function for
assessing goodness of fit and determining best fit quantities and
confidence levels.  In Table 2, we give the best-fit values and 90\%
confidence level statistical uncertainties for the spectral parameters
and elemental abundances for both the full equilibration model and the
model assuming energy exchange between electron and ions through
Coulomb collisions.

\par From a statistical point of view, the reduced $\chi^2$ values given in
Table 2 are formally unacceptable. However, systematic errors in the
experimental data due to calibration uncertainties, for example, have
not been included, nor have we included uncertainties due to the
atomic physics or the incompleteness of the X-ray line lists used in
the spectral models.  This last point is particularly important since the
pattern of residuals for all the SNRs show significant excesses at photon
energies of $\sim$0.9 keV and $\sim$1.2 keV which are due to missing Fe
L-shell lines in the models.  As a representation of the \asca\ data, 
however, the Sedov model does a considerably better job than the
phenomenological NEI model discussed above (\S\ref{phenomenological}),
which was specifically introduced in order to merely reproduce the
observed \asca\ spectra with little astrophysical motivation. Although
the phenomenological model requires a larger number of free
parameters, the Sedov model provides fits with comparable or even
better quality. Given our present state of knowledge of the
instrumental calibration and the atomic physics of X-ray spectral line
emission, we believe that the Sedov model describes the \asca\ spectra
about as well as can be expected.

\par

Our fitted column densities from the \asca\ spectra agree quite well
with estimates of the total column toward each remnant based on the
\ion{H}{1} gas in the Galaxy (Heiles \& Cleary 1979) and in the LMC
(McGee \& Milton 1966). In particular, the total column of \ion{H}{1}
gas is greatest toward N49, N63A, and N49B (values of $\sim$$2.1\times
10^{21}$ atoms cm$^{-2}$) and is least toward N23, DEM71, N132D, and
0453$-$68.5 (values of $\sim$$1.2\times 10^{21}$ atoms cm$^{-2}$),
which correlates quite well with the X-ray results. The quantity of
absorbing gas in the Galaxy along the line-of-sight to the LMC sets
the minimum column density we should expect to see. This is $(5 -
6)\times 10^{20}$ atoms cm$^{-2}$, which appears to be consistent with
the columns derived for DEM71, N23, and perhaps 0453$-$68.5.

\par

The reader will notice that both the fitted shock temperature and
ionization timescale are consistently higher for the Coulomb
equilibration timescales model compared to the full equilibration one,
as a direct consequence of the lag of the electron temperature with
respect to the ion temperature in the former case. However, the
difference in temperature is small, $\lsim$30\%, and comparable to the
statistical uncertainty in the parameters.  The ionization timescale
differs by a much larger amount, up to factors of $\sim$3, the
consequences of which we discuss below when we compare the ionization
and dynamical ages.  The other fitted parameters, i.e., column
density, normalization and the elemental abundances, also agree to
within the statistical errors. We note that the quality of the fits as
indicated by the reduced $\chi^2$ values is nearly the same under the
two different assumptions we have made about electron-ion
equilibration timescales, which is unfortunate since it means that our
study can not use this to discriminate between the two cases.

\par  From the fitted normalization of the X-ray spectra ($N$) and shock
temperature ($T_s$), we estimate the ISM number density of hydrogen
($n_{0}$), the Sedov dynamical age ($t_{d}$), the initial explosion
energy ($E_{0}$), and the swept-up mass ($M_{SU}$), using the
following equations:

\begin{equation}
n_{0} = 13 \left(\frac{N}{10^{12}{\rm\, cm}^{-5}}\right)^{1/2}  
\left(\frac{\theta_{R}}{{\rm 10^{\prime\prime}}}\right)^{-3/2} 
\left(\frac{D}{50{\rm\, kpc}}\right)^{-1/2} 
\left(\frac{\Omega}{4\pi}\right)^{-1/2}  {\rm H\ cm}^{-3},
\end{equation}
\begin{equation}
t_{d} = 10^3 \left(\frac{kT_{s}}{{\rm 1\, keV}}\right)^{-1/2} 
\left(\frac{\theta_{R}}{{\rm 10^{\prime\prime}}}\right) 
\left(\frac{D}{50{\rm\, kpc}}\right) {\rm yr},
\end{equation}
\begin{equation}
E_{0} = 0.33\times 10^{51} \left(\frac{kT_{s}}{{\rm 1\, keV}}\right) 
\left(\frac{N}{10^{12}{\rm\, cm}^{-5}}\right)^{1/2} 
\left(\frac{\theta_{R}}{{\rm 10^{\prime\prime}}}\right)^{3/2} 
\left(\frac{D}{50{\rm\, kpc}}\right)^{5/2} 
\left(\frac{\Omega}{4\pi}\right)^{-1/2} {\rm erg},
\end{equation}
\begin{equation}
M_{SU} = 26 \left(\frac{N}{10^{12}{\rm\, cm}^{-5}}\right)^{1/2} 
\left(\frac{\theta_{R}}{{\rm 10^{\prime\prime}}}\right)^{3/2} 
\left(\frac{D}{50{\rm\, kpc}}\right)^{5/2} 
\left(\frac{\Omega}{4\pi}\right)^{1/2} M_\odot,
\end{equation}
where $\Omega$ is the solid angle subtended by the remnant shell with
respect to the remnant center, and is a factor that is intended to
take account of possible incompleteness of the spherical structure.
Spherical symmetry ($\Omega = 4 \pi$) was assumed in order to compute
the values of the derived quantities given in Table 2.  The physical
radii in Table 2 are based on the angular X-ray size from Mathewson et
al.\ (1983) and our assumed distance of 50 kpc to the LMC.  In
calculating the swept-up mass we assumed that the mean ionic weight
was $1.4\,m_{\rm H}$.

Figure \ref{fig2} plots the ambient ISM hydrogen number density and derived
explosion energy versus radius for the several remnants.  The ambient
density is uncorrelated with remnant size for our sample. There are
some potential biases which work against finding SNRs in the upper
right corner of this figure (radiative cooling) or the lower left
(rapid expansion), but given the lack of a statistically significant
correlation, these issues are moot. 

The range of derived $E_0$ values vary over an order of magnitude from
a high of $6 \times 10^{51}$ erg for N132D to a low of $5 \times
10^{50}$ erg for N23. Recall that $E_0$ depends on $\Omega^{-1/2}$ and
that we assumed the maximum value for $\Omega$ ($=4\pi$, i.e.,
spherical symmetry), so the explosion energies we quote are in some
sense lower bounds. In particular, N23, which has the lowest derived
$E_0$ value, also is the remnant that has the most asymmetric X-ray
image. Uncertainty in the distance to individual SNRs due to the
thickness of the LMC, for example, is probably not much more than
10\%, which translates to a $\sim$25\% error in $E_0$.  As for the
ambient density, we find no statistically significant evidence for a
dependence of the derived explosion energy on remnant size.  This is
in contrast to optical studies of LMC SNRs (Dopita 1979, Danziger \&
Leibowitz 1985) that have claimed a rather strong dependence of $E_0$
on remnant size, even for some of the same remnants as in our sample.
These results were never particularly convincing (at least to us)
since they were based on extrapolating the pressure behind the shock,
as estimated from the density-sensitive [\ion{S}{2}] lines, to
estimate the energy density in the interior of the entire remnant.
Blair, Kirshner, \& Chevalier (1981) have pointed out that the
apparent relation between $E_0$ and radius could be explained if the
[\ion{S}{2}] emitting regions are dominated by magnetic pressure,
rather than thermal pressure as assumed otherwise.  Our explosion
energy values are based on the well constrained observables of shock
temperature, radius, and emission measure, and rely on the
applicability of the Sedov evolutionary model.  Whether or not this is
justified is the next issue we consider.

One of the assumptions of Sedov-phase remnant models is that they are
adiabatic, or energy conserving, which implies that radiative cooling
is insignificant. Cox (1972) identified a number of evolutionary
phases in the life of an SNR as it makes the transition from an
adiabatic remnant to a fully radiative one.  In his picture, the onset
of radiative cooling occurs when the temperature distribution begins
to deviate from the adiabatic model (temperature sag phase) at an age
given by
\begin{equation}
t_{sag} = 1.33\times 10^4 \left(\frac{E_0}{10^{51} {\rm\, ergs}}\right)^{2/11} 
\left(\frac{\Lambda}{10^{-22} {\rm\, ergs\, cm^3\, s^{-1}}}\right)^{-5/11} 
\left(\frac{n_0}{1{\rm\, cm^{-3}}}\right)^{-7/11} \rm yr,
\end{equation}
where the cooling coefficient $\Lambda$ is of order $10^{-22} \rm\,
ergs\, cm^3\, s^{-1}$.  Using $E_0$ and $n_0$ from Table 2, $t_{sag}$
varies from $7\times 10^3$ yr for N63A to $3\times 10^4$ yr for
0453$-$68.5. For all the remnants in our sample, $t_{sag}$ is at least
50\% larger than the dynamical age quoted in Table 2 showing that, at
least from this point of view, the Sedov model is justified.

As we mentioned in \S\ref{sedovmodel}, we can estimate the ionization
age ($t_{i}$) of our remnant sample from the fitted $n_{0}t_{i}$
values, by employing the $n_{0}$ value derived from the emission
measure (eqn. 1). Because of the difference in fitted ionization
timescales between the full and Coulomb temperature equilibration models, 
there is a significant difference in the inferred ionization ages as 
well. In Figure \ref{fig3} we plot the correlation between ionization
age (for both models) and dynamical age. For all remnants, the full
equilibration model predicts an ionization age that is considerably
lower than the dynamical age with differences ranging from
$\sim$1.3$\sigma$ (for 0453$-$68.5) to $\sim$9$\sigma$ (for N49B). On
the other hand, under the assumption of Coulomb equilibration
timescales there is excellent agreement (a difference of
$\lsim$1.2$\sigma$) between the ionization and dynamical ages for N23,
N49, DEM71, and 0453$-$68.5 although the ionization ages of N132D,
N63A, and N49B all continue to remain low. This discrepancy for the
latter three remnants cannot be
resolved by invoking clumping of the X-ray emitting gas, since that
would increase our estimate for $n_{0}$ and consequently lower the
inferred ionization ages even further. Similarly, reducing $\Omega$
from its spherically symmetric value of $4\pi$ would also result in
reduction of the ionization ages as well.  We believe that the most
likely reason for the discrepancy is that the Sedov model has
overestimated the dynamical ages of these remnants.

In this case, it is not coincidental that the three remnants with
significant differences between their ionization and dynamical ages
are also the remnants with unusually high inferred values of initial
SN explosion energy: $\gsim$$3\times 10^{51}$ erg. A plausible
explanation for both discrepancies is that these remnants exploded
within preexisiting low-density cavities in the ISM.  The low interior
density allows the SN blast wave to propagate rapidly to the cavity
wall, where it then encounters denser gas, begins to slow down, and
emits copious amounts of X-rays.  This scenario naturally results in a
reduction of both the dynamical age and the initial explosion energy
from that inferred assuming the Sedov model. Note that Hughes (1987)
proposed precisely this same model to account for the difference
between the Sedov dynamical age and the kinematic age determined from
the expansion of high velocity oxygen-rich filaments in N132D (more on
this in \S\ref{n132d} below).  It is encouraging that our independent
estimate of N132D's age based on its ionization timescale confirms
this earlier result and it gives us confidence in the identification
of N63A and N49B as two more examples of this phenomenon in the LMC.
This turns out to be extremely important, since it allows us to
identify these three remnants as the core-collapse supernovae of
massive stars whose strong stellar winds modified the surrounding ISM.

The four other SNRs, N23, N49, DEM71, and 0453$-$68.5, are fully
consistent with the Sedov model under the assumption that the
electrons and ions do not attain the same temperature at the shock
front, but rather exchange energy slowly through Coulomb
collisions. This hypothesis appears to be supported by other
observations of SNRs; perhaps the best evidence to date comes from the
remnant of SN1006 in the Galaxy.  Based on comparing shock models to
the observed far-UV emission lines of the remnant, Laming et al.\
(1996) find that the electrons at the shock front in SN1006 can attain
a temperature of no more than $\sim$20\% of the ion temperature and
that even lower equilibration values are prefered by their data. This
is the strongest observational evidence in the literature against
full electron-ion equilibration at SN shock fronts.  We close this
section by noting that the consistency of these four SNRs with the
Sedov model appears to extend even to the value we find for their
average initial explosion energy $E_0 = (1.1\pm0.5)\times 10^{51}$
erg, where the uncertainty comes from the RMS object-to-object
scatter.

\section{Abundances}
\subsection{Correlations}
\label{correlations}

Figure \ref{fig4} displays the complete set of abundance results for the seven
LMC remnants, where we have chosen to average the fitted abundances
(and errors) that come from our two different assumptions about the
timescale for electron-ion temperature equilibration. Overall, we find
abundances that are less than cosmic by factors of 2 to 5,
as expected for interstellar gas in the LMC. At first glance the
different remnants yield rather similar abundances for the same
elemental species, but a closer look reveals some notable exceptions
that we discuss further below.

In order to check for systematic effects that might have biased the
fitted abundances we carried out the following correlation analysis.
First we determined the mean metallicity for each remnant by
calculating the error-weighted average of the fitted abundance values
(presented in the bottom half of Table 2), including both sets of
values from the full and Coulomb equilibration models.  We searched
for correlations between these mean metallicities and the various
fitted or derived parameters using the Spearman rank-order test.  The
correlation coefficents ($r_s$) for mean metallicity versus $N_{\rm
H}$, $T_s$, $n_0 t_i$, and $N_s$ are $-0.14$, 0.04, 0.61, and 0.21,
respectively, although none of the correlations were statistically
significant (among these the highest significance was $\sim$93\% for
metallicity versus $n_0 t_i$). This is very encouraging since it shows
that there are no obvious systematic biases in the derived abundances
that are directly linked to the spectral fits.  Likewise there are
no statistically significant correlations between mean metallicity and
any of the derived parameters $E_0$, $n_0$, and $M_{SU}$.  There are,
however, two statistically significant correlations: mean metallicity
versus age ($r_s = -0.89$, significance $\sim$99.6\%) and and mean
metallicity versus radius ($r_s = -0.96$, significance $\sim$99.97\%).
Plots of these latter two correlations, plus the correlation of
metallicity versus swept-up mass, are shown in Figure \ref{fig5}. The error
bars displayed are estimates based on the weighted RMS scatter from
the individual abundance measurements of the six species considered.

The correlation in the left panel of Figure \ref{fig5} shows that the
smaller remnants are somewhat enhanced in metallicity with respect to
the larger ones.  In fact, this correlation also holds when the
elemental species are considered individually.  Specifically, the
oxygen, neon, silicon, and iron abundances show correlations with
radius that are significant at the 99.97\%, 94\%, 96\% and 95\%
confidence levels, respectively, while the magnesium and sulfur
abundances are essentially uncorrelated with remnant size.  Note that
these correlations can be seen in Figure \ref{fig4} since the data
points corresponding to the individual remnants are plotted for each
species from left to right in order of remnant size.

\hii\ regions in the LMC are known to show a significant
correlation between distance from the galaxy center and their oxygen
abundance based on measurements of [\ion{O}{2}] and [\ion{O}{3}] from
warm photoioinized gas. The abundance gradient has a value of $-0.048
\pm 0.019$ dex/kpc (Kobulnicky 1997). We determined the deprojected
radial distances of our remnants from the center of the LMC assuming a
flat disk geometry, an inclination angle of 27$^\circ$ between the
normal to the disk and the line-of-sight, position angle of
165$^\circ$, and center at 5$^{\rm h}$24$^{\rm m}$,
$-$69$^\circ$48$^\prime$. The remnants in our sample cover
galactocentric radial distances from 0.3 kpc to 3.3 kpc, which would
suggest a 30\% difference in abundance based on the \hii\ regions.
In fact, the mean metallicity is uncorrelated with distance from the
center of the LMC and our data are best described by a model with no
gradient in chemical abundance with galactocentric distance.

The interpretation we favor is that the metallicity-radius correlation
indicates the presence of newly produced SN ejecta in at least some of
the remnants of the sample.  This is not particularly surprising,
since we know that small, young Galactic remnants like Tycho and Cas A
show very strong and prominent X-ray emission from supernova ejecta,
while larger ones, like the Cygnus Loop and Vela, don't.  Furthermore,
earlier work by us (Hughes et al.\ 1995) on the \asca\ X-ray spectra
of the three LMC SNRs N103B, 0509$-$67.5, and 0519$-$69.0 (with radii
between 3 and 4 pc) revealed them to be dominanted by ejecta. It is
perhaps a little surprising that the correlation is so good with
radius or age but not with swept-up mass. Apropos to this point is the
observation that the only remnant out of these seven previously known
to contain newly minted SN ejecta is N132D, which has the largest
swept-up mass in our sample!  Clearly the fate of the metals ejected
by a SN is complex, which makes prediction of their emission
properties as a function of time rather difficult.  Such a study is
beyond the scope of our work here and we direct the interested reader
to Tenorio-Tagle (1996), where a discussion of some of the relevant
issues for type II supernovae is given.

Figure \ref{fig5} puts the level of enrichment (i.e., the difference between
the average metallicity of the smaller SNRs compared to the larger
ones) at a value of $\sim$0.1. Interestingly, this difference is
comparable to the amount of metals that models predict should be
produced and ejected by certain mass ranges of progenitor stars.  For
example if we take the nucleosynthetic yields for a progenitor with a
main-sequence mass of 13 $M_\odot$ from Thielemann, Nomoto, \&
Hashimoto (1996) and dilute it with 200 $M_\odot$ of swept-up material
(the median value in our sample), then the fractional abundances of O,
Ne, and Mg would increase by $\sim$0.1. In the same way the abundances
of Si, S, and Fe would increase by an amount equal to $\sim$0.4.  The
nucleosynthetic yields of the lighter elements, in particular, are a
strong function of main sequence mass and, thus, diluting the ejecta
from a 25 $M_\odot$ progenitor with 200 $M_\odot$ of ISM would result
in the O abundance increasing by $\sim$1.  The preceding points are
intended to indicate the order of magnitude agreement between the
amount of enrichment we actually see in the LMC SNR sample and the
amount that might be expected in the very simplest scenario imaginable
for the mixing of ejecta and ISM in SNRs.

In addition to this statistical evidence for ejecta enrichment, there
is direct evidence as well.  In particular, we refer the reader to
Figure \ref{fig4} where the anomalously high iron abundance for DEM71 can be
clearly seen. Although some other individual abundances are slightly
discrepant with respect to their neighbors (e.g., the magnesium
abundance for N49B), only the case of the iron abundance of DEM71 is
statistically significant. This will be discussed in more detail below
(\S\ref{dem71}).

\subsection{Depleted Abundances?}
\label{depletion}

Another question concerning our abundance measurements that must be
addressed is the possible depletion of metals onto dust grains in the
interstellar medium. Interstellar gas is believed to be significantly
depleted in magnesium, silicon, calcium, and iron (with abundances
roughly an order of magnitude less than solar), but only slightly
depleted in oxygen and sulfur.  Depletions depend on the mean gas
density of the environment, in general, growing larger with increasing
gas density (Mathis 1990).  An important constraint is that neon and
argon, as inert, noble gases, are not expected to form grains and thus
should not be depleted in the gas-phase.  Our abundance results appear
to be in conflict with this picture of the dusty ISM.  In fact, the
abundance we find for neon is essentially the same as that of
magnesium, silicon, and sulfur, while oxygen and iron are no more than
about a factor of two less.

Vancura et al.~(1994) study the effects of interstellar grains on the
predicted X-ray and IR emission of supernova shock waves and find that
the X-ray spectra of SNRs are substantially influenced by the gradual
destruction of dust grains, primarily through collisions with protons
and He nuclei, in the post-shock flow.  Their models are based on a
planar shock geometry and do not include the deceleration of the blast
wave, so their results are not directly applicable to our
measurements.  Nevertheless, with this as a warning we attempt a
comparison. The fraction of initial mass remaining in grains is a
strong function of the swept-up column density behind the shock. As an
estimate of the swept-up column for our remnants we use the quantity
$n_0 R/6$, which simply assumes that the entire mass of shocked gas in
the SNR is contained in a uniform density shell of thickness $R/12$
and takes the average value of the swept-up column at the inner and
outer edges of the shell. The smallest value of this quantity among
the remnants in our sample is $2.3\times 10^{18}$ cm$^{-2}$ (for
0453$-$68.5) while the largest is $1.7\times 10^{19}$ cm$^{-2}$ (for
N63A).  According to Figure 3 in Vancura et al.~(1994) the fraction of
initial mass that should have remained in grains is 60\% for our
minimum swept-up column and 20\% for the maximum one.  For the species
magnesium, silicon, and iron, which are supposed to be nearly entirely
depleted in the gas-phase, then the mean abundances should vary by
about a factor of 2.4 over this range in swept-up column.  We just
don't see such large variations in the abundances of these elements as
a function of swept-up column density. Furthermore, there is not even
a significant correlation between swept-up column and abundance for
magnesium and iron, although the correlation for the silcon abundance
versus swept-up column density is significant at the 97.4\% confidence
level.  Weighing all the evidence presented in this and the preceding
paragraph leads us to conclude that we have not detected evidence for
the destruction of swept-up interstellar dust grains on the expected
timescales in these LMC SNRs.  This suggests that either the gas-phase
of the ISM is not significantly depleted in metals or the timescale
for the destruction of dust grains behind SN shock waves is shorter
than previously thought.

\subsection{Metal Abundances of the LMC}
\label{lmc_abun}

We determine the gas-phase abundances of the LMC by averaging, for
each elemental species, the fitted abundances from the seven
remnants. As before we include both sets of values corresponding to
the full and Coulomb equilibration models. The uncertainties quoted
come from the weighted RMS scatter among the 14 measurements.  For
sulfur, the values from 0453$-$68.5 and N49B are not included, since
these SNRs provided only upper limits.  Table 3 lists the average
abundances and uncertainties expressed as $[M/H] = 12 + \log ({\rm M/H})$
and Figure \ref{fig6} plots the abundances, relative to the cosmic
values we assumed, versus  elemental species.

\par

The third column of Table 3 lists the LMC chemical abundances of
oxygen, neon, and sulfur determined from UV/optical spectra of \hii\ 
regions as summarized in a review by Dufour (1984). The
agreement between the X-ray SNRs and \hii\ regions for the
different elements is impressive. More recently, RD92 presented the
results of a large study to establish a consistent set of global
abundances for the Large and Small Magellanic Clouds covering the
elements from $Z=2$ to $Z=63$. This work was based on spectral
analyses of F supergiants for the abundances of the heavier elements
(e.g., Mg, Si, and elements with $Z>18$ ) (Russell \& Bessell 1989)
and the modelling of spectra from \hii\ regions and SNRs for the
lower $Z$ elements (e.g., He, O, Ne, and S) (Russell \&
Dopita 1990). Our values and the ones from RD92 are also in general
agreement with the possible exception of magnesium and silicon, where
the RD92 values appear anomalously high compared to the other species,
although it should be noted that the abundance of silicon was quoted
in RD92 as being highly uncertain. It is now clear that these species
are not anomalously abundant in the ISM of the LMC compared to, say,
neon and sulfur.  Thus the results from X-ray SNRs have proved to be
especially valuable in filling this important gap in our
knowledge of the abundances of the intermediate mass elements
and in providing a set of accurate abundance values determined in a
consistent manner for elemental species over a broad range of atomic
number.

\par

Now a few statements about the reliability of our abundance estimates.
The abundances of neon, magnesium, silicon, and sulfur that we derive
from the \asca\ X-ray spectra of SNRs are least subject to systematic
uncertainty since they come from fits to K-shell emission lines of
well understood helium-like and hydrogen-like ionic species.  Our iron
abundance is based on the L-shell blend at $\sim$1 keV, so there is
some systematic uncertainty due to inadequate or poorly known atomic
physics and the incompleteness of the line lists used in the spectral
models. Our oxygen abundance is also somewhat less reliable because
the oxygen K-shell lines are near the low energy cutoff of the SIS
detectors where the efficiency is low and the calibration is less
certain.  Furthermore, of all the species considered, the oxygen
abundance is most sensitive to the precise variation of temperature
and ionization through the interior of our Sedov models.  For example,
over the outer 10\% radius of the Sedov models (from which nearly all
of the X-ray emission comes), the ionization fraction of O VIII
(hydrogen-like) varies by more than an order of magnitude (decreasing
from the edge in toward the center). By contract, Si XIV (also
hydrogen-like) is the dominant ionic species and varies by less than a
factor of 2 over the same radial range.  Thus, our models predict that
the oxygen emission should be coming from a narrow shell near the edge
of the SNR, while the silicon emission should be distributed over a
larger radial extent.  Future higher spatial resolution X-ray
observations with \axaf\ will be able to measure these variations and
thereby provide more accurate abundance results that are less
dependent on the model details. However, the comparison between our
results and previous ones shown in Figure \ref{fig6} suggests that the
systematic error in the oxygen and iron abundances derived from our
work is not much more than $\sim$ 0.1--0.2, with our values being
slightly lower than the optical/UV ones.

\section{Limits on High Energy X-Ray Emission}

X-ray imaging at energies above $\sim$4 keV has been one of the most
important recent developments in the search for pulsar-powered
synchrotron nebulae (PSN) in supernova remnants.  Studies with \asca\
in the past 2 years have confirmed the existence of, or directly
detected, PSN in several Galactic SNRs including Kes 73 (Blanton \&
Helfand 1996), G11.2$-$0.3 (Vasisht et al.~1996), W44 (Harrus, Hughes,
\& Helfand 1996), CTA-1 (Slane et al.~1997), and MSH 11-6{\sl 2}
(Harrus, Hughes, \& Slane 1998). The reason for this is simple: above
4 keV thermal emission from shock-heated gas is negligible, while hard
power-law emission from a PSN remains relatively strong. The discovery
of a PSN in a supernova remnant is pivotal, since it conclusively
establishes the remnant as the product of a massive-star core-collapse
SNe. The only LMC SNRs that can, without question, be put in this
class are the SNRs 0540$-$69.3 and N157B both of which contain rapidly
rotating pulsars with periods of 50.3 ms (Seward, Harnden, \& Helfand
1984) and 16.1 ms (Marshall et al.~1998), respectively, and whose
X-ray emissions are dominated by the nonthermal power-law from the
PSN.  Because of the known distance to the LMC, these PSN have
well-determined soft (0.15--4.5 keV) X-ray luminosities of $9.5\times
10^{36}\,\rm erg\, s^{-1}$ (0540$-$69.3) and $2.5\times 10^{36}\,\rm
erg\, s^{-1}$ (N157B).

\par

A factor of $>$2500 separates the luminosity of 0540$-$69.3 from that
of the faintest known Galactic PSN which surrounds PSR B1853+01 in
W44.  It is quite possible that PSN with luminosities in this range
are present in some of the LMC SNRs in our sample even though all are
clearly dominated by thermal emission. However the arcmin imaging
capability of \asca, which is adequate for spatially resolving thermal
and nonthermal emission in Galactic SNRs, is insufficient for doing so
for LMC SNRs, where a 2 pc diameter PSN would subtend merely
8$^{\prime\prime}$.  Thus any information on PSN from \asca\ will have
to come from the integrated spectra of individual LMC remnants.  At
this time we do not claim unambiguous detection of nonthermal X-ray
emission for any of the 7 SNRs in this study, because the SIS emission
observed above $\sim$3 keV is comparable to the fluctuations in
background.

\par

In order to set quantitative limits, we fitted the high energy portion
of the SIS spectra to a power-law model ($dN/dE \sim E^{-\alpha_p}$)
with a fixed photon index of $\alpha_p = 2.0$.  For the four fainter
SNRs (N23, DEM71, 0453$-$68.5, and N49B) the band fitted was 3--10
keV, while for the brighter ones (N49, N63A, and N132D) we used
the 4--10 keV band.  Acceptable fits were obtained in all cases.  The
3-$\sigma$ upper limits to the flux of this component were then used
to calculate unabsorbed luminosities in the soft X-ray band as
shown in Table 4.

\par

These luminosity values are in the range expected for PSN (Seward \&
Wang 1988). The high limit for N63A is rather interesting since this
is one of the remnants, along with N132D and N49B, that is likely to
have arisen from a core-collapse supernova and therefore might be
expected to contain a PSN.  Note that the high upper limit value for
N132D largely reflects the short \asca\ exposure on the remnant.  The
limits we set for the others are consistent with them {\it not}
containing rapidly spinning young pulsars, in agreement with our
age estimates assuming Sedov evolution.

\par

Current theories predict that Type Ia SNe should not produce pulsars,
so their remnants should not contain PSN. It is possible that these
SNRs may show nonthermal X-ray emission from electrons accelerated to
high energies at the shock front by the first-order Fermi process as
observed in SN1006 (Koyama et al.~1995). This emission tends to be
steeper spectrum ($\alpha_p = -2.4$ to $-3$) and fainter (SN1006's
soft $L_x$ is $\sim$$4\times 10^{34}$ erg s$^{-1}$) than that of PSN
and therefore much more difficult to detect in LMC SNRs. The upper
limit to the high energy luminosity for DEM71, which is the only
securely identified remnant of a Type Ia SN in the sample, is about an
order of magnitude higher than SN1006's luminosity.

\section{Comments on Individual Remnants} 

\subsection{N23}
\label{n23}

N23 is not a particularly well studied remnant.  Chu \& Kennicutt
(1988, hereafter CK88) tentatively classify it as the remnant of a
massive star SNe (Pop I) based on the high density of OB stars nearby
and its proximity to a modest molecular cloud (Cohen et al.\ 1988).
Recently Banas et al.\ (1997) searched for CO emission on smaller
spatial scales ($\sim$20$^{\prime\prime}$) than the Cohen survey, but
found nothing significant that could be associated with the remnant.
We note that N23 has one of the lowest column densities in the sample
(a factor of 3--4 less than N49 and N49B), which may make direct
association with a molecular cloud unlikely.  As the smallest remnant
in our sample, N23 is an important object that supports the
statistical evidence for SN ejecta enrichment discussed above.

\subsection{N49}

Images of N49 in the optical, X-ray, and radio bands (Vancura et
al.~1992; Dickel et al.~1995) are in general agreement as far as the
broad picture of its morphology is concerned.  The remnant is
brightest in the southeast and trails off in brightness toward the
northwest.  On smaller scales, however, individual features differ in
relative brightness across the electromagnetic spectrum.  The
asymmetry and high X-ray and optical flux are due to the enhanced
ambient density toward the southeast, where a moderate-sized molecular
cloud sits (Banas et al.~1997). The large X-ray absorbing column
density we derive, further indicates a high density environment. This
is also true of the mean optical reddening of N49, $E(B-V) \sim 0.37$
(Vancura et al.~1992), which is quite a bit more than the reddening
due to the Galaxy, $E(B-V) \sim 0.07$ (Fitzpatrick 1985).  We note
that the X-ray absorbing column density and interstellar reddening
toward N49 are in excellent agreement with the ratio of these
quantities, $N_{\rm H}/E(B-V) = (6.8\pm 1.6) \times 10^{21}$ atoms
cm$^{-2}$ mag$^{-1}$, given by Ryter, Cesarsky, \& Audouze (1975).

By comparing shock models to their optical and UV spectra of N49,
Vancura et al.~(1992) estimate the abundances of several species.  For
oxygen and sulfur, their abundances are some 50\% higher than ours,
while their iron abundance is in numerical agreement with ours. The
other species in common (neon, magnesium, and silicon) agree to within
a factor of two, but the optical results are based on lines from only
a single ion and are therefore potentially less reliable.  Since
Vancura et al.~(1992) quote no abundance uncertainties it is difficult
to assess the significance of these differences.

Vancura et al.~(1992) also develop a self-consistent scenario for the
evolution of N49.  Optical emission comes from slow shocks (velocity
range of 40--270 \kms) driven into dense clouds (20--940 cm$^{-3}$)
distributed in thin sheets that are located at places where the SN
blast wave is interacting with the nearby molecular cloud. X-ray
emission comes from a fast shock (730 \kms) propagating through the
intercloud medium (preshock density of 0.90 cm$^{-3}$).  They derive a
Sedov age of roughly 5400 yr by assuming an explosion energy of
$10^{51}$ ergs.  In this picture the prevailing ISM was largely
unaffected by the progenitor star, in direct contrast to the view
espoused by Shull et al.~(1985), where the SN blast wave is
interacting with the dense shell of the progenitor star's Str\"omgen
sphere. 

Our results show that the X-ray data on N49 are consistent with Sedov
evolution in two ways. First, the initial SN explosion energy
we derive from our fitted shock temperature is consistent with the
canonical value of $10^{51}$ ergs.  Second the two ages we determine
for the remnant, i.e., the ionization age and the Sedov dynamical age,
are entirely consistent with each other. Thus the X-ray, optical, and UV
data all tend to support the Vancura et al.~(1992) picture for the
evolution of N49.

There is at least one other interesting aspect of N49, namely that it
is positionally coincident with the 5 March 1979 $\gamma$-ray burst
event, one of only three known soft $\gamma$-ray repeaters (SGR).
Rothschild et al.~(1994) claim that the burst position of SGR0525$-$66
is coincident with an unresolved source at the northern edge of N49,
based on soft X-ray images from the \rosat\ high resolution imager
(HRI).  Our \asca\ results do not preclude the possibility that this
source has a hard power-law spectrum. In fact, the upper limit we
derive for the power-law emission from N49 (Table 4) corresponds to an
HRI count rate of 0.022 s$^{-1}$, which is greater than the actual
count rate, $0.0151\pm0.0013$ s$^{-1}$, of the unresolved source in
the HRI image.  In addition, our upper limit to the hard power-law
emission implies a luminosity $<$$4\times 10^{35}$ ergs s$^{-1}$
(2--10 keV band), which is comparable to the persistent X-ray
luminosity of AX1805.7$-$2025 ($3\times 10^{35}$ ergs s$^{-1}$;
Murakami et al.~1994).  This source is the counterpart to one of the
other SGRs (SGR1806$-$20) and is believed to be an isolated neutron
star powering a radio nebula in the SNR G10.0$-$0.3 (Kulkarni et
al.~1994). Getting back to N49, Dickel et al.~(1995) claim to have
found no evidence for a radio or optical counterpart to SGR0525$-$66.
They do note that a compact radio feature lies near the quoted
position of the unresolved X-ray source, but they dismiss it as being
unrelated largely due to a $5^{\prime\prime}$ difference in the radio
and X-ray positions. However this conclusion may be premature, since a
few arcseconds is roughly the accuracy for absolute position
determination by \rosat.  It is probably fair to say that the issue of
the counterpart to SGR0525$-$66 has not been completely resolved and
awaits further study with \axaf.

\subsection{N63A}
\label{n63a}

N63A is the second brightest X-ray emitting SNR in the LMC and its
derived shock temperature, ambient density, and abundances (Table 2)
turn out to be very similar to those of the other bright LMC SNRs, N49
and N132D.  However, unlike both N49 and N132D, N63A's morphology
varies quite a bit across the electromagnetic spectrum.  At optical
wavelengths the remnant consists of three bright lobes of emission
embedded in the extended diffuse \hii\ region N63. Two of the three
lobes consist of shock-heated gas while the third is a photoionized
region.  The shock-heated gas appears to be of normal composition for
the LMC and is expanding at moderate velocities, $\sim$250 km s$^{-1}$
(Shull 1983). In the radio band N63A shows a thick asymmetric
shell-like structure $\sim$65$^{\prime\prime}$ in diameter with
enhancements at the positions of the bright optical features (Dickel
et al.~1993). The X-ray image (Mathewson et al.~1983) also displays a
thick shell of roughly the same size as the radio one but without any
obvious enhancements corresponding to features at the other
wavelengths.  Curiously, the bright optical region is only about \onethird\
the size of the X-ray or radio shells and is displaced westward by
about 15$^{\prime\prime}$ from their respective centers. 

Based on its location in an \hii\ region, as well as its positional
coincidence with the OB association NGC 2030, it is not surprising
that this remnant is considered to have had a Pop I progenitor (CK88).
This could also be consistent with our suggestion that N63A exploded
within a wind-blown cavity in the ISM.  We derive the highest ambient
ISM density for this remnant, nearly 4 cm$^{-3}$, yet other signs of
such a high density environment are not obvious. For example, unlike
both N49 and N132D, there is no evidence for molecular emission near
N63A from either the Columbia survey of the LMC (Cohen et al.~1988) or
from higher angular resolution SEST data (Israel et al.~1993). It is
clear that a lot more work will be needed before we obtain a
comprehensive picture of this enigmatic SNR and its environment.

\subsection{DEM71}
\label{dem71}

There is a class of SNRs that show only hydrogen lines in their
optical spectra with virtually no emission from collisionally excited
forbidden lines, e.g., [\ion{O}{3}] and [\ion{S}{2}], as usually seen
from SNRs.  Tycho's remnant and SN1006, believed to be from Type Ia
SNe, are examples of such so-called Balmer-dominated SNRs in the
Galaxy. There are also four such remnants in the LMC (Tuohy et
al.~1982).  Two of them, 0509$-$67.5, and 0519$-$69.0, show strong
X-ray emission lines of the metals Si, S, Ar, Ca, and Fe (Hughes et
al.~1995), which are known to be characteristic of Type Ia supernova
ejecta.  The young inferred ages of these SNRs ($\lsim$ 1500 yr) are
consistent with the ejecta-dominated nature of their X-ray emission.
DEM71 is another LMC Balmer-dominated SNR that is older and more
evolved than the two just discussed. Based on its age, size, and the
estimated amount of gas swept-up by the blast wave (see Table 2),
DEM71 is believed to be in a transition state in which the X-ray
emission is changing from being dominated by SN ejecta to being
dominated by swept-up interstellar gas. (The fourth LMC example of
this class, 0548$-$70.4, is a much weaker X-ray source and is not in
the current \asca\ sample, so it will not be discussed further.)

The Balmer-dominated optical spectra can be interpreted rather well in
the context of a model in which a high-velocity interstellar shock
overtakes at least partially neutral interstellar gas (Chevalier \&
Raymond 1978). When examined in detail the H$\alpha$ line from these
remnants shows a distinctive two-component line profile, consisting of
narrow and broad components.  The width of the broad component
provides a nearly direct measurement of the shock velocity.  For
DEM71, Smith et al.~(1991) determined a probable range for the shock
velocity of 300--800 km s$^{-1}$, corresponding to shock temperatures
of 0.11--0.76 keV.  This range is in very good agreeement with our
X-ray-derived value of $kT_s \sim 0.8$ keV.

\par

It is now reasonably secure that Balmer-dominated SNRs are the
remnants of Type Ia SNe. In light of this point it is very interesting
that DEM71 is the only remnant in the current \asca\ sample that shows
direct evidence for enhanced abundances. As clearly shown in Figure
\ref{fig4} the iron abundance of DEM71 is about a factor of 2 higher
than all the others in the sample. This corresponds to an amount of
``extra'' iron present in DEM71 of roughly $M_{Fe} \sim 0.06\,
M_\odot$, which is about 10\% of the total amount of iron believed to
be ejected by a Ia supernovae (Nomoto, Thielemann, \& Yokoi
1984). This discovery provides further support for a causal connection
between Balmer-dominated SNRs and Type Ia SNe and it highlights the
important of such objects for understanding the chemical enrichment of
the ISM.  Follow-up observations with \axaf\ and searches for coronal
[\ion{Fe}{14}] $\lambda$5303 emission are planned in order to
determine the spatial distribution of the excess iron in DEM71 and to
search for other elemental species that might be enhanced.

\subsection{N132D}
\label{n132d}

N132D is the one of the brightest LMC remnants and is probably the
best studied of them all.  Its popularity with optical observers is
due in part to the fact that it contains high-velocity oxygen-rich
ejecta that identifies the remnant as having come from the
core-collapse SN of a massive star.  This ejecta lies near the
geometric center of the remnant and is surrounded by a larger
limb-brightened shell of material with normal LMC abundances (see
references in Morse, Winkler, \& Kirshner 1995). These two components
also appear, more or less, in the radio (Dickel \& Milne 1995) and the
X-ray (Hughes 1987) images of N132D, although individual features when
examined in detail do vary with wavelength. It is not known if the
X-ray emitting features near the center of the remnant are
oxygen-rich.  One clue, suggesting that they may not be enhanced, was
first presented by Blair, Raymond, \& Long (1994) who noted that the
X-ray peaks near the center were generally associated with
low-velocity, normal composition, optical filaments, rather than with
high-velocity oxygen-rich ones. \hst\ observations of N132D (Morse et
al.~1996) provide a clear confirmation of this effect.

The high-velocity oxygen-rich filaments provide an independent
estimate of the age of N132D.  The most detailed analysis of the
velocity field of N132D has been carried out by Morse et al.~(1995)
who find an expansion velocity of $\sim$1650 \kms\ for filaments with
a mean radius of 5.3 pc, that results in a kinematic age of $\sim$3200
yr assuming undecelerated motion. This value is considerably less than
the Sedov dynamical age we quote in Table 2.  This discrepancy between
the kinematic age and the Sedov dynamical age was noted before (Hughes
1987) and it led to the suggestion that N132D exploded within a
low-density cavity evacuated by the stellar wind and ionizing
radiation of the high-mass progenitor. In this scenario, the outer
shell of emission is the main blast wave that has reached the edge of
the cavity and is now interacting with the high density material
there.  The high density of the ambient medium can be explained by the
presence of a giant molecular cloud associated with N132D (Banas et
al.~1997). Our work with the \asca\ spectra provides yet another
independent estimate of the age, based on the ionization timescale,
which is also considerably less than the Sedov dynamical age and
supports the cavity scenario.  

Our derived N132D abundances for the elements from oxygen to iron are
higher by, at most, a factor of three than those derived by Hwang et
al.~(1993) using data from the \EO.  There are several reasons for
this discrepancy.  First our \asca\ SIS data are of considerably
higher quality (in all respects: higher sensitivity, lower background,
better spectral resolution, more accurate calibration, and coverage of
a broader X-ray band) than the \einstein\ SSS data used by Hwang et
al.  Our Sedov models provide a more realistic description of the
physical conditions in the SNR than the single-temperature
single-timescale phenomenological NEI models used previously.  The
fitted column density in our work is about a factor of 2 more than
that of Hwang et al., which tends to increase our fitted abundances
relative to theirs.  Note that both our fitted column density and
Hwang et al.'s are consistent with the total column of neutral
hydrogen toward N132D, $N_{\rm HI} \approx 1.4\times 10^{21}$
cm$^{-2}$.  We believe our X-ray determination of the absorbing column
density is more accurate because it is based on a more realistic
spectral model and because the \asca\ data are better calibrated in
the important band below 1 keV.  Additionally, we can point to the
excellent correlation we get between the \ion{H}{1} column and the X-ray
column for the seven SNRs in our sample to support the general
accuracy of our results.

Another determination of N132D's abundances was recently presented by
Favata et al.~(1997) using X-ray spectral data from BeppoSAX.  These
results are in strong disagreement with the results we present here
and thus the discrepancy requires some explanation.  First we note
that the column density derived by Favata et al., $N_{\rm H} \sim
3\times 10^{21}$ cm$^{-2}$ (but with large error), is considerably
larger than the total column of neutral hydrogen toward N132D quoted
above.  Because of their higher column density their abundances will
be spuriously high. Their two-component NEI model is a somewhat better
model than the single-component model used by Hwang et al.~(1993), but
it is not as realistic as our Sedov-phase model.  However, the
difference in spectral models does not fully explain the difference in
derived abundances.  We also applied a two-component NEI model to the
\asca\ SIS data of N132D and found best-fit temperatures of 0.5--0.6
keV (component A) and 2.2--3.5 keV (component B) with a ratio of
emission measures of 0.05--0.06 (B/A), values that are similar to
Favata et al.  However, this two-component NEI model still resulted in
low values for the abundances using the \asca\ SIS data: $0.31\pm
0.05$ for oxygen, $0.43\pm 0.06$ for neon, and $0.10\pm 0.02$ for
iron.  An oxygen abundance as high as 1.2 (as derived from the
BeppoSAX data) would produce a huge K$\alpha$ oxygen line in the SIS
that we just don't see. We believe that the main difference in our
results arises from the poorer spectral resolution of the BeppoSAX
data compared to the data from the SIS. The abundances are determined
from the emission line strengths, so clearly better spectral
resolution is important. However, the derived abundances are sensitive
to the ionization timescales which can only be constrained accurately
if lines from helium- and hydrogen-like ionic species can be resolved.
We resolve these lines in the SIS data, while the BeppoSAX data does
not.

We also have an astrophysical objection to the large abundances quoted
by Favata et al.  If their abundances were correct, it would require
that we are seeing all of the enriched ejecta from a massive
progenitor.  In particular, according to the Favata et al.\ results, the
amount of ``extra'' mass in oxygen above the amount in the swept-up
ISM is $\sim$4 $M_\odot$ which is comparable to the entire mass of
oxygen ejected by a 25 $M_\odot$ progenitor (Thielemann, Nomoto, \&
Hashimoto 1996). Since we see oxygen-rich filaments in the optical
band, at least some of the ejecta have cooled below $\sim$$10^4$ K and
therefore cannot be contributing to the X-ray emission.  Clearly 
definitive answers to questions about the amount and distribution of
enriched X-ray--emitting ejecta in N132D must await \axaf\ observations.

\subsection{0453$-$68.5}

This remnant is located in the western edge of the LMC and, like N23
above, is also rather poorly studied. It has been tentatively
classified as the remnant of a Type Ia SNe (Pop II) by CK88
due to the lack of nearby OB stars, \hii\
regions, and CO emission.  This interpretation is consistent with the
low preshock density we derive for 0453$-$68.5 ($n_0 = 0.30$
cm$^{-3}$, the lowest value in the sample), although a lack of
evidence for an overabundance of iron, as seen for DEM71, is a (weak)
countervailing factor.  The line-of-sight column density is low,
similar to the values found for N23 and DEM71, and probably arises
mostly from the interstellar medium of the Galaxy.

\subsection{N49B}
\label{n49b}

Optically, N49B consists of two sets of bright knots separated on the
sky by approximately $1\myarcmin.4$ and embedded in a patchy network
of diffuse nebulosity. Its X-ray appearance is entirely different,
showing a somewhat limb-brightened, nearly-circular region of emission
that encompasses both sets of optical knots.  On the basis of its
location within the \hii\ region DEM 181, CK88
identify N49B with a Pop I progenitor, a picture that is consistent
with our finding that N49B is the result of an explosion within a
preexisting cavity in the ISM. The relatively high column density
toward this remnant suggests that its association with molecular
emission (Cohen et al.~1988) may be more than an accidental projection
effect.

\section{Summary}

We have carried out a systematic analysis of the \asca\ SIS X-ray
spectra of seven luminous SNRs in the LMC.  Our work has resulted in
the following conclusions.

\begin{description}

\item[(1)] The spectral data are described well by a NEI model that
includes the time evolution of density and temperature given by the
Sedov similarity solution for the dynamical evolution of SNRs.
This model was a better description of the data than any single or
multiple component phenomenological NEI model that we tried.

\item[(2)] The X-ray data provide 4 independent constraints on the 3
independent parameters of the Sedov model. The observational
constraints are: radius, shock temperature, emission measure, and
ionization timescale.  The parameters of the Sedov model are age,
initial explosion energy, and ambient ISM density.  There is an
additional ambiguity in the models that arises from the unknown
relationship between the energy imparted to electrons and ions at the
shock front.  We model two extreme situations: full equilibration (in
which the electron and ion temperatures are assumed to be equal) and
Coulomb equilibration (in which the electrons gain energy from the
ions though Coulomb collisions only).

\item[(3)] Four of the remnants in the sample, N23, N49, DEM71, and
0453$-$68.5, are fully consistent with Sedov evolution under the
assumption of Coulomb equilibration. For this group of remnants the
mean explosion energy is $(1.1\pm 0.5)\times 10^{51}$ ergs in
agreement with the canonical value. The key to discriminating between
the full and Coulomb equilibration models comes from comparing the
Sedov dynamical age with the age determined from the ionization
timescale.

\item[(4)] When the derived parameters of the other three SNRs, N63A,
N132D, and N49B, are examined in detail, they appear to be internally
inconsistent. In particular the age determined by the ionization
timescale is considerably less than the Sedov dynamical age,
regardless of whether one assumes the full or Coulomb equilibration
models.  Furthermore the initial SN explosion energies are large
$\gsim$$3\times 10^{51}$ ergs. Both of these discrepancies can be
explained by invoking a scenario in which the remnants exploded within
preexisting cavities or bubbles in the ISM and that the X-ray emission
we see now comes from the blast wave interacting with the dense
material at the cavity wall.  This was previously suggested for N132D
to explain the discrepancy between this remnant's Sedov age and the
kinematic age determined from the expansion of high-velocity
oxygen-rich filaments. Our independent estimate of N132D's age from
its ionization timescale confirms this earlier result and it gives
confidence that N63A and N49B are actually two more examples of this
phenomenon.  This result is not particularly surprising since it is
precisely the manner in which massive stars are predicted to modify
their environment.

\item[(5)] We find statistical evidence for enrichment by supernova
ejecta in the sense that smaller remnants show a somewhat higher mean
metallicity than the larger ones.  On the other hand, the mean
metallicity does not correlate with swept-up mass.  In the case of the
Balmer-dominated SNR DEM71, which is likely to be the remnant of a
Type Ia supernova, the derived abundance of iron is about a factor of
two larger than the other remnants in the sample. This corresponds to
an excess of iron over the amount in the ISM of $\sim$0.06 $M_\odot$,
which is about 10\% of the total amount of iron ejected by a Type Ia
SN.

\item[(6)] All things being considered, however, the middle-aged,
evolved SNRs in our sample are in general dominated by swept-up ISM and
so can be used to estimate the mean LMC gas-phase abundances. We find
that the common elements from oxygen to iron have abundances 0.2--0.4
times solar, consistent with previous results based on optical and UV
data, but without the anomalous overabundance of Mg and Si seen by
others. This work demonstrates the validity of using X-ray spectra of
SNRs to measure the current chemical composition of interstellar gas
and it has the potential to provide a significant new constraint on
the chemical evolution and star formation history of the Cloud.

\item[(7)] We set limits on the hard ($\gsim$3 keV) X-ray flux of the
remnants that correspond to luminosities of from $1.9\times
10^{35}\rm\, ergs\, s^{-1}$ to $2.2\times 10^{36}\rm\, ergs\, s^{-1}$
in the 0.2--4.0 keV band. These limits indicate that the remnants in
the sample do not contain very luminous pulsar-powered synchrotron
nebulae, as is consistent with our picture of them as evolved
middle-aged SNRs. However, the quoted luminosity range is
consistent with the possiblity that they contain weaker synchrotron
nebulae with pulsars losing energy at a slower rate than is the case
in the Crab Nebula, for example. The discovery of such pulsar-powered
synchrotron nebulae would definitively identify the progenitor as a
massive star that underwent a core collapse SN.

\end{description}

\acknowledgments

We gratefully acknowledge the numerous scientists, engineers, and
other members of the team that designed, developed, and built \asca.
We thank T.~Murakami for allowing us early access to the \asca\ data
on N49 and D.~Helfand for useful discussions and comments on the
manuscript.  This research has made use of the NASA/IPAC Extragalactic
Database (NED) which is operated by the Jet Propulsion Laboratory,
California Institute of Technology, under contract with the National
Aeronautics and Space Administration.  This research was partially
supported by NASA grants NAG5-2684 and NAG5-0638 and  NSF grant
INT-9308299 to JPH.

\begin{table}[p]
\begin{center}
TABLE 1\\
{{\it ASCA} Observation Log of LMC SNRs} \\
~\\
\begin{tabular}{l l l l l} \hline \hline
		&Other  $^{\rm a}$ 	&Date of 		&Exposure
 &Source Count Rate $^{\rm b}$ 		\\ 
SNR	&Name		&Observation		&Time (ksec)	&SIS0/SIS1 (count\ s$^{-1}$)  	\\ \hline	
0453$-$68.5	&		&95 November 25		&38.7		& 0.14/0.11	\\
0505$-$67.9	&DEM71		&95 October 8		&28.7		& 0.50/0.39	\\
0506$-$68.0	&N23		&93 November 9		&32.9		& 0.35/0.29	\\
0525$-$66.0	&N49B		&94 October 10		&37.4		& 0.35/0.25	\\ 
0525$-$66.1	&N49		&94 October 10		&37.4		& 0.59/0.55	\\
0525$-$69.6	&N132D		&93 September 23	&5.2		& 4.07/3.39	\\
0535$-$66.0	&N63A		&93 November 21		&18.0		& 3.16/2.62	\\	\hline \hline		
\multicolumn{5}{l}{$^{\rm a}$ N: Henize (1956); DEM: Davies, Elliott,
\& Meaburn (1976)} \\
\multicolumn{5}{l}{$^{\rm b}$ From within the source extraction regions.} \\

\end{tabular}
\end{center}

\end{table}
\normalsize

\def\tightenB{\def\baselinestretch{1.2}}

\tightenB

\small
\begin{table}[p]
\begin{center}
TABLE 2\\
{Sedov Model Fits to LMC SNRs $^{\rm a}$} \\
\end{center}
\begin{tabular}{l c c c c c c c c c} \hline \hline
	&$R$ 	&$N_{\rm H}$		&$kT_{s}$ 	&log\,$n_{0}t_{i}$	&$N_S$		&$n_{0}$ 	&$t_{S}$	&$E_{0}$ 	&$M_{\rm SU}$ \\ 
SNR	&(pc)	&(10$^{21}\,$cm$^{-2}$)	&(keV)	&(cm$^{-3}\,$s)		&$(10^{12}\, {\rm cm}^{-5}$)	&(H\,cm$^{-3}$)	&(10$^{3}$\,yr)	&(10$^{51}$\,erg)	&($M_{\odot})$ \\ \hline
N23		&6.7	&0.8(0.3)		&0.53(0.10)	&10.89(0.27)		&0.33(0.12)		&1.6		&3.8		&0.46		&70	\\
		&	&0.7(0.3)		&0.59(0.15)	&11.16(0.21)		&0.35(0.10)		&1.7		&3.6		&0.53		&70	\\
N49		&8.2	&2.2(0.3)		&0.58(0.05)	&11.33(0.18)		&1.59(0.28)		&2.6		&4.4		&1.5		&210	\\
		&	&2.1(0.3)		&0.62(0.05)	&11.52(0.13)		&1.57(0.24)		&2.6		&4.3		&1.6		&200	\\
N63A		&8.5	&1.4(0.1)		&0.62(0.04)	&11.23(0.17)		&3.77(0.55)		&3.9		&4.5		&2.6		&330	\\
		&	&1.4(0.2)		&0.68(0.07)	&11.38(0.13)		&3.95(0.49)		&3.9		&4.2		&2.9		&340	\\
DEM71		&10.4	&0.4(0.3)		&0.82(0.14)	&10.58(0.11)		&0.21(0.06)		&0.67		&4.7		&1.1		&110	\\
		&	&0.6(0.3)		&0.83(0.41)	&11.03(0.10)		&0.35(0.13)		&0.86		&4.7		&1.4		&140	\\
N132D		&12.1	&1.3(0.2)		&0.68(0.06)	&10.95(0.16)		&4.70(0.79)		&2.5		&6.1		&5.4		&630	\\
		&	&1.3(0.1)		&0.78(0.14)	&11.19(0.07)		&5.43(0.27)		&2.7		&5.7		&6.7		&680	\\
0453$-$68.5	&15.0	&$<$ 1.1		&0.51(0.14)	&10.68(0.30)		&0.12(0.07)		&0.30		&8.7		&0.91		&140	\\
		&	&$<$ 0.5		&0.70(0.39)	&11.01(0.34)		&0.11(0.05)		&0.28		&7.4		&1.2		&130	\\
N49B		&17.0	&2.6(0.4)		&0.41(0.06)	&10.78(0.11)		&1.15(0.57)		&0.75		&10.9		&2.7		&520	\\ 
		&	&2.5(0.4)		&0.44(0.10)	&11.08(0.16)		&1.37(0.55)		&0.82		&10.6		&3.1		&560	\\
		&	&			&	  &			&			&		&		&		&	\\
\end{tabular}
\begin{tabular}{l c c c c c c c c} \hline
		&		&	 	&		&		&		&		&		&$L_{\rm X}$[0.5\,--\,5\,keV]	\\ 
SNR		&O		&Ne	 	&Mg		&Si
&S		&Fe		&$\chi^2_\nu$/d.o.f.		&(10$^{36}$\,erg\,s$^{-1}$) \\ \hline
N23		&0.27(0.06)	&0.47(0.09)	&0.53(0.13)	&0.38(0.11)	&0.50(0.34)	&0.31(0.08)	&1.4/172	&2.5 	\\
		&0.32(0.08)	&0.45(0.11)	&0.50(0.13)	&0.36(0.10)	&0.52(0.32)	&0.28(0.07)	&1.3/172	& 	\\
N49		&0.32(0.04)	&0.48(0.05)	&0.44(0.09)	&0.45(0.05)	&0.61(0.10)	&0.29(0.05)	&2.4/247	&6.3	\\
		&0.38(0.04)	&0.48(0.09)	&0.43(0.09)	&0.43(0.07)	&0.58(0.12)	&0.29(0.05)	&2.5/247	&	\\
N63A		&0.25(0.05)	&0.50(0.05)	&0.50(0.06)	&0.40(0.05)	&0.29(0.07)	&0.31(0.04)	&2.8/146	&20	\\
		&0.28(0.05)	&0.45(0.08)	&0.46(0.08)	&0.36(0.05)	&0.27(0.09)	&0.30(0.04)	&2.7/146	&	\\
DEM71		&0.22(0.03)	&0.55(0.08)	&0.50(0.12)	&0.26(0.09)	&0.47(0.27)	&0.62(0.12)	&2.2/178	&3.4	\\
		&0.27(0.05)	&0.48(0.08)	&0.39(0.11) 	&0.24(0.08)	&0.57(0.26)	&0.47(0.08)	&2.1/178	&	\\
N132D		&0.21(0.03)	&0.44(0.06)	&0.46(0.07)	&0.34(0.05)	&0.36(0.11)	&0.30(0.04)	&1.8/229	&30	\\
		&0.24(0.02)	&0.42(0.04)	&0.41(0.04)	&0.31(0.03)	&0.36(0.09)	&0.27(0.02)	&1.8/229	&	\\
0453$-$68.5	&0.21(0.05)	&0.28(0.07)	&0.33(0.13)	&0.23(0.14)	&$<$ 0.78	&0.26(0.08)	&2.1/121	&0.98	\\
		&0.26(0.09)	&0.25(0.09)	&0.33(0.15)	&0.23(0.14)	&$<$ 0.71	&0.28(0.10)	&2.1/121	&	\\
N49B		&0.22(0.03)	&0.42(0.05)	&0.74(0.10)	&0.28(0.06)	&$<$ 0.30	&0.21(0.04)	&2.5/169	&3.2	\\ 
		&0.23(0.05)	&0.36(0.07)	&0.63(0.12)	&0.25(0.07)	&$<$ 0.39	&0.16(0.03)	&2.4/169	&	\\ \hline \hline
\end{tabular}

\begin{flushleft}
$^{\rm a}$ For each remnant, the results of fits with models assuming
full electron and ion temperature equilibration are given in the upper
line, while those from models assuming Coulomb equilibration
timescales are given in the lower line. The distance to the LMC was
taken to be 50 kpc.  Values in parentheses represent the 90\%
confidence statistical errors.

\end{flushleft}

\end{table}

\normalsize

\begin{table}[p]
\begin{center}
TABLE 3\\
{LMC Abundances} $^{\rm a}$ \\
~\\
\begin{tabular}{l c c c} \hline \hline
	& \asca\ X-ray		& \hii          & \hii\ regions, SNRs, \\
Species & SNRs $^{\rm b}$	& regions $^{\rm c}$ & supergiant
stars $^{\rm e}$   \\ \hline
$[$O/H$]$   & $8.21\pm0.07$	&$8.43\pm0.08$	& $8.35\pm0.06$ \\
$[$Ne/H$]$  & $7.55\pm0.08$	&$7.64\pm0.10$	& $7.61\pm0.05$ \\
$[$Mg/H$]$  & $7.08\pm0.07$	&$\ldots$	& $7.47\pm0.13$ \\
$[$Si/H$]$  & $7.04\pm0.08$	&$\ldots$	& $7.81\,^{\rm e}\phantom{\pm0.13}$ \\ 
$[$S/H$]$   & $6.77\pm0.13$	&$6.85\pm0.11$	& $6.70\pm0.09$ \\
$[$Fe/H$]$  & $7.01\pm0.11$	&$\ldots$	& $7.23\pm0.14$ \\ \hline \hline
\multicolumn{4}{l}{$^{\rm a}$ Expressed as $[{\rm M/H}]\equiv 12 + \log ({\rm M/H})$}\\
\multicolumn{4}{l}{$^{\rm b}$ This work, errors are the RMS variation among
the seven SNRs} \\
\multicolumn{4}{l}{$^{\rm c}$ Dufour (1984)} \\
\multicolumn{4}{l}{$^{\rm d}$ Russell \& Dopita (1992)} \\
\multicolumn{4}{l}{$^{\rm e}$ Highly uncertain} \\

\end{tabular}
\end{center}

\end{table}
\normalsize

\begin{table}[p]
\begin{center}
TABLE 4\\
{Upper Limits to Power-Law X-ray Emission} \\
~\\
\begin{tabular}{l c c} \hline \hline
	& Energy Band	& 3-$\sigma$ Upper Limit \\
	& Fitted 	& $L_X$ (0.2--4 keV) \\
SNR	& (keV)		& ($10^{35}\rm\, ergs\, s^{-1}$) \\ \hline
N23		& 3--10		& $\phantom{0}2.6$ \\
N49		& 4--10		& $\phantom{0}7.6$ \\
N63A		& 4--10 	& $13.5$	   \\
DEM71		& 3--10		& $\phantom{0}5.1$ \\
N132D		& 4--10		& $22.4$	   \\
0453$-$68.5	& 3--10         & $\phantom{0}2.3$ \\
N49B		& 3--10         & $\phantom{0}1.9$ \\ \hline \hline
\end{tabular}
\end{center}

\end{table}

\normalsize

\clearpage

\figcaption[fig1.ps]{(a)-(g) \asca\ SIS spectra of seven LMC SNRs. In each
panel the data and 1-$\sigma$ statistical errors are shown by the
crosses. The solid histogram is the best-fit spectrum from the Sedov
model assuming full electron-ion temperature equilibration. For clarity of
display only, the SIS0 and SIS1 data and corresponding models were
averaged. The actual spectral analysis was done by jointly fitting to both
sets of SIS data. \label{fig1}}

\figcaption[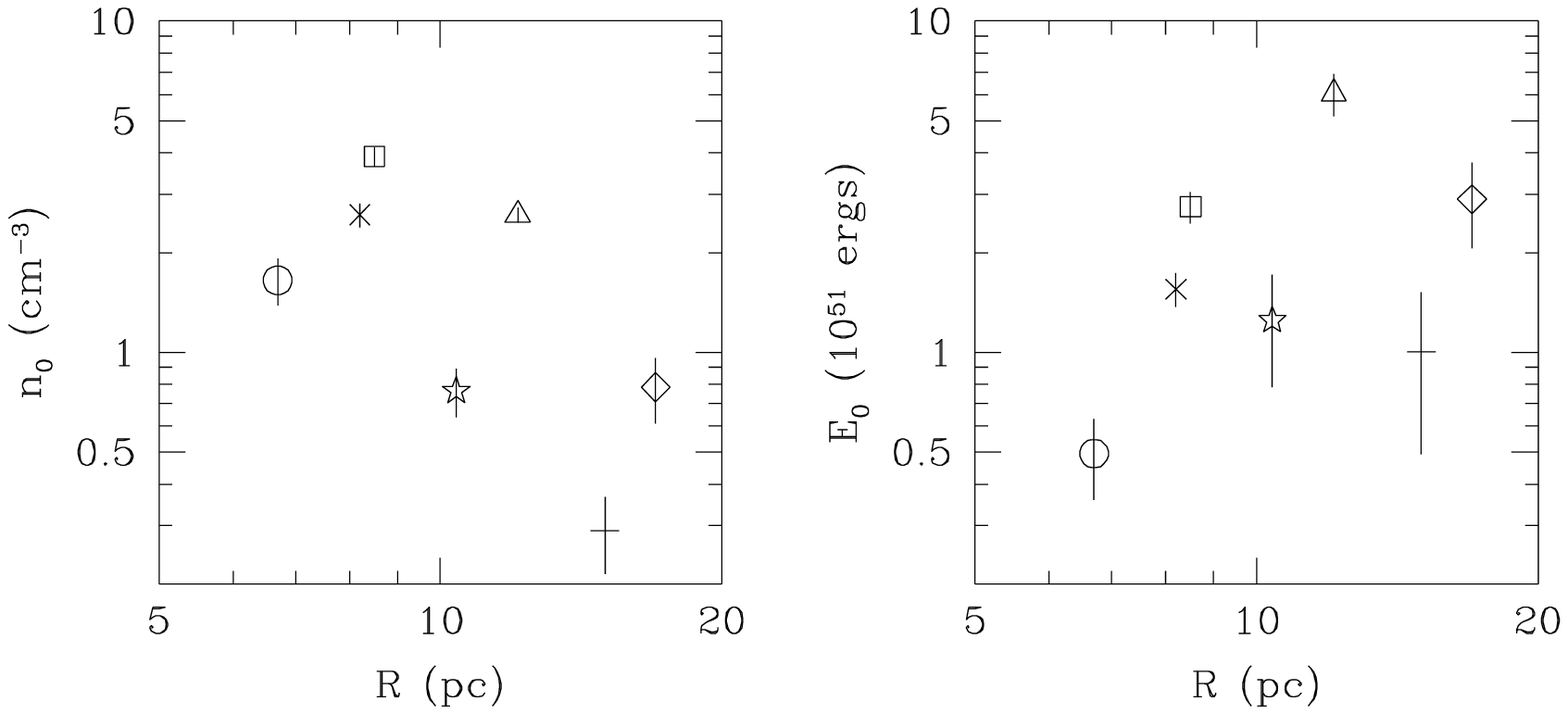] {Ambient ISM hydrogen number density versus
remnant radius (left panel) and initial supernova explosion energy
versus radius (right panel). Values for all seven remnants are shown
using different symbols: the circle, cross, square, star, triangle,
plus sign, and diamond correspond to N23, N49, N63A, DEM71, N132D,
0453$-$68.5, and N49B, respectively. Error bars include only the
statistical uncertainty (at the 90\% confidence level) from fits for
temperature and emission measure. Results plotted are the average of
the values obtained under the two different model assumptions about
the timescale for electron-ion temperature equilibration.\label{fig2}}

\figcaption[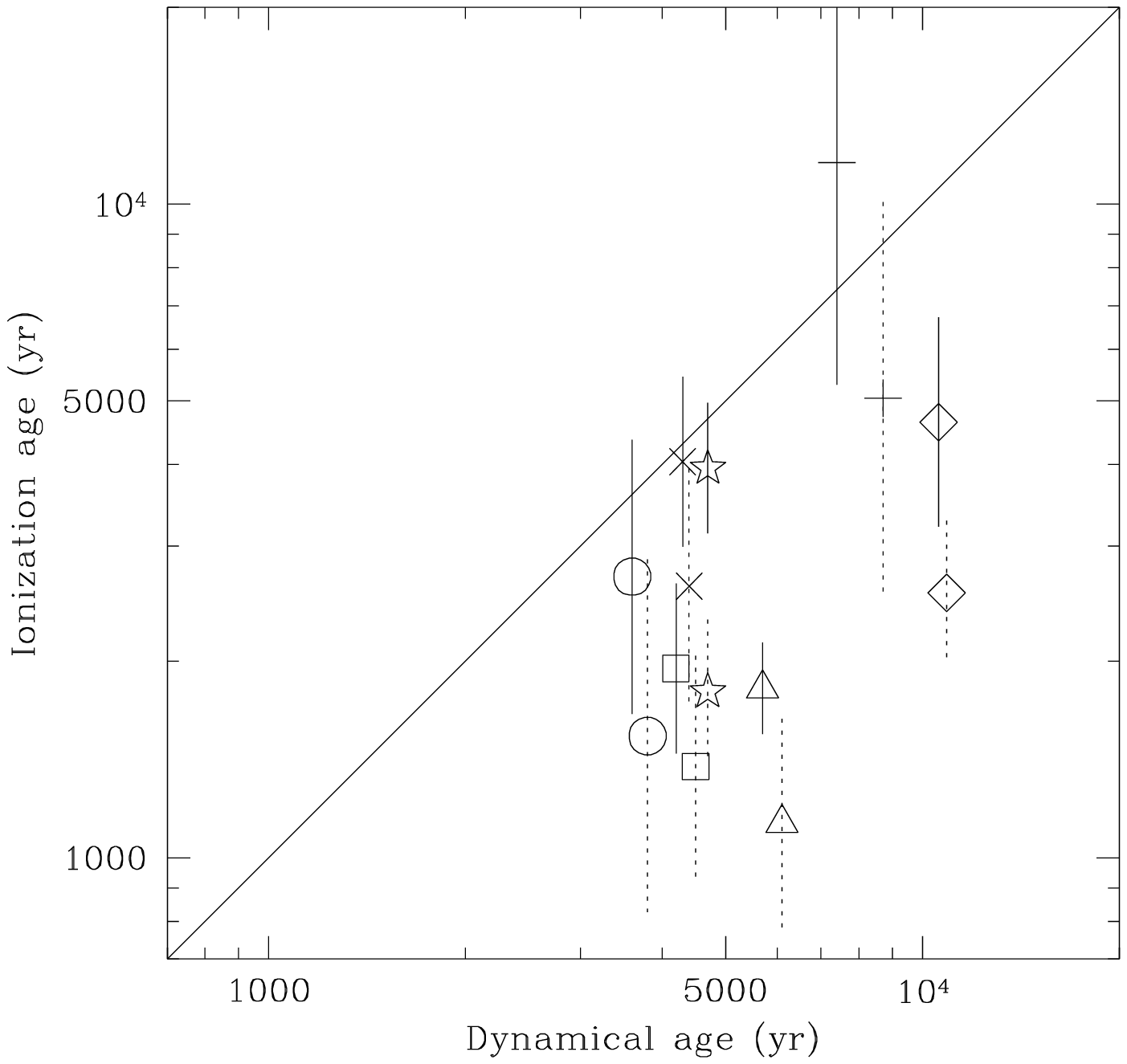] {Remnant age determined from the fitted
ionization timescale versus the dynamical age derived from the Sedov
solution. The different symbol types refer to the different remnants
(circle, cross, square, star, triangle, plus sign, and diamond
correspond to N23, N49, N63A, DEM71, N132D, 0453$-$68.5, and N49B,
respectively). Error bars include only the statistical uncertainty
from fits for the ionization timescale.  Data points with dotted error
bars (at the 90\% confidence level) are from the results assuming
full electron-ion temperature equilibration at the SNR shock front,
while the solid error bars refer to results using Coulomb equilibration 
timescales.\label{fig3}}

\figcaption[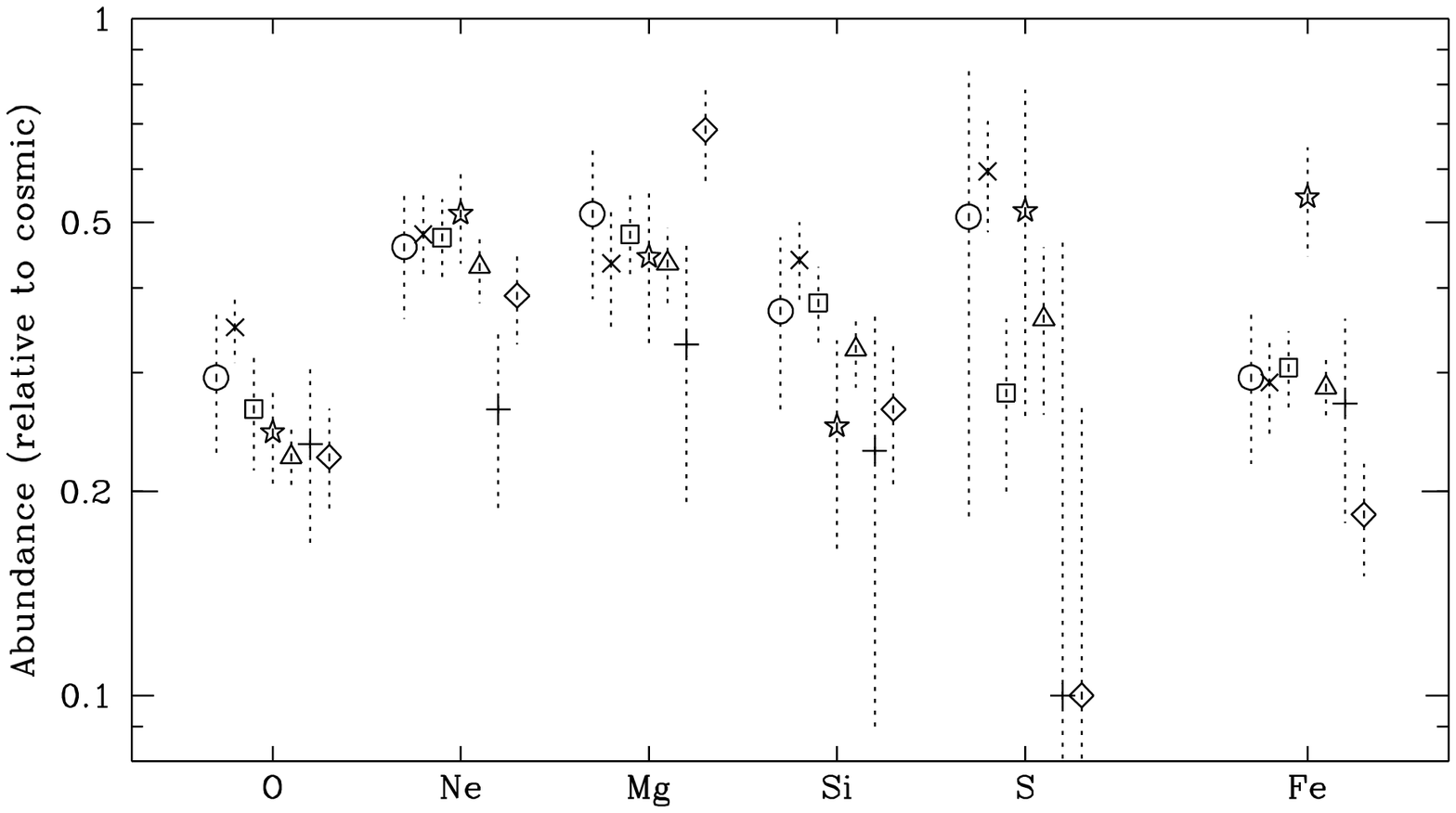] {Elemental abundances relative to the cosmic values
derived from \asca\ X-ray spectroscopy of LMC SNRs. Values for all
seven remnants are shown using different symbols: the circle, cross,
square, star, triangle, plus sign, and diamond correspond to N23, N49,
N63A, DEM71, N132D, 0453$-$68.5, and N49B, respectively. Error bars
include only the statistical uncertainty (at the 90\% confidence
level) from the spectral fits.  Results plotted are the average of the
values obtained under the two different model assumptions about the
timescale for electron-ion temperature equilibration. We can only 
determine upper
limits to the S abundances of 0453$-$68.5 and N49B. These points are
plotted at a value of 0.1 and the upper limit lies at the top of the
error bar.\label{fig4}}

\figcaption[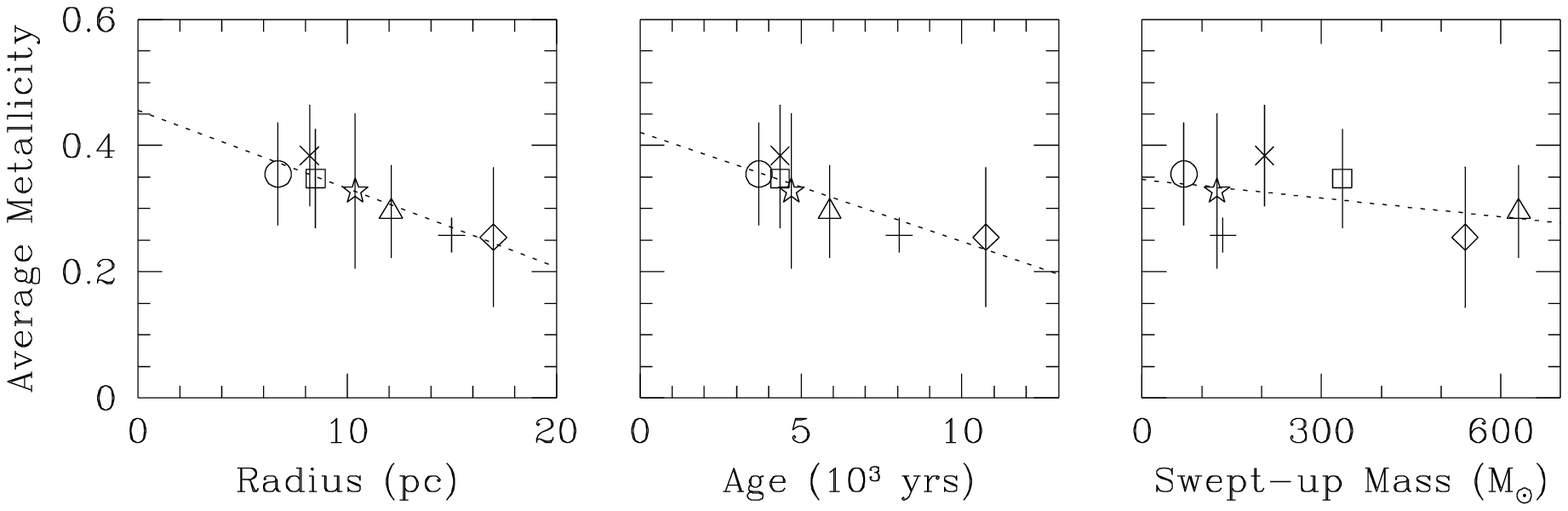] {Mean metallicity (relative to cosmic values)
versus radius (left panel), dynamical age derived from the Sedov
solution (middle panel), and swept-up mass (right panel).  The
different symbol types refer to the different remnants (circle, cross,
square, star, triangle, plus sign, and diamond correspond to N23, N49,
N63A, DEM71, N132D, 0453$-$68.5, and N49B, respectively). Error bars
represent the RMS scatter in the individual fitted abundances of the
various elemental species considered. Results plotted are the average
of the values obtained under the two different model assumptions about
the timescale for electron-ion temperature equilibration. The only 
significant correlations are between metallicity and either radius or
age.\label{fig5}}

\figcaption[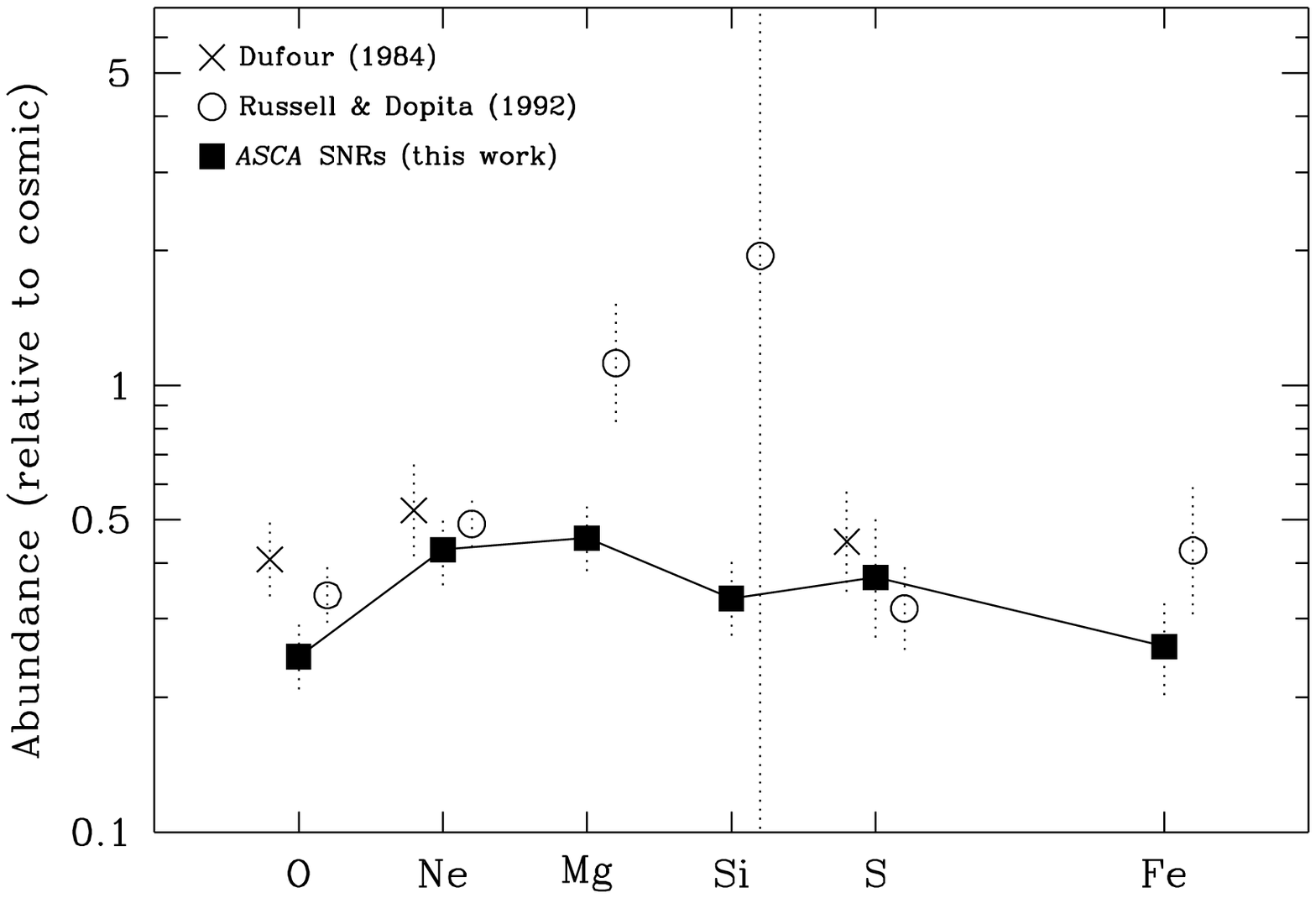] {Average chemical abundances of the LMC relative
to the cosmic values plotted as a function of elemental species. The
gas phase abundances of the LMC derived from optical and UV studies of
\hii\ regions are shown as the crosses (Dufour 1984).  More
recent results from Russell \& Dopita (1992), shown as the circles,
are based on detailed modelling of optical/UV spectra from \hii\
regions, SNRs, and supergiant stars. The silicon abundance in Russell
\& Dopita's work is highly uncertain as indicated by our use of a
large error bar for this species.  The filled squares, showing our
results from \asca\ X-ray spectroscopy of SNRs, are in excellent
agreement with the values derived from longer wavelength
observations. \label{fig6}} 

\clearpage

\clearpage
\begin{figure}
\plotfiddle{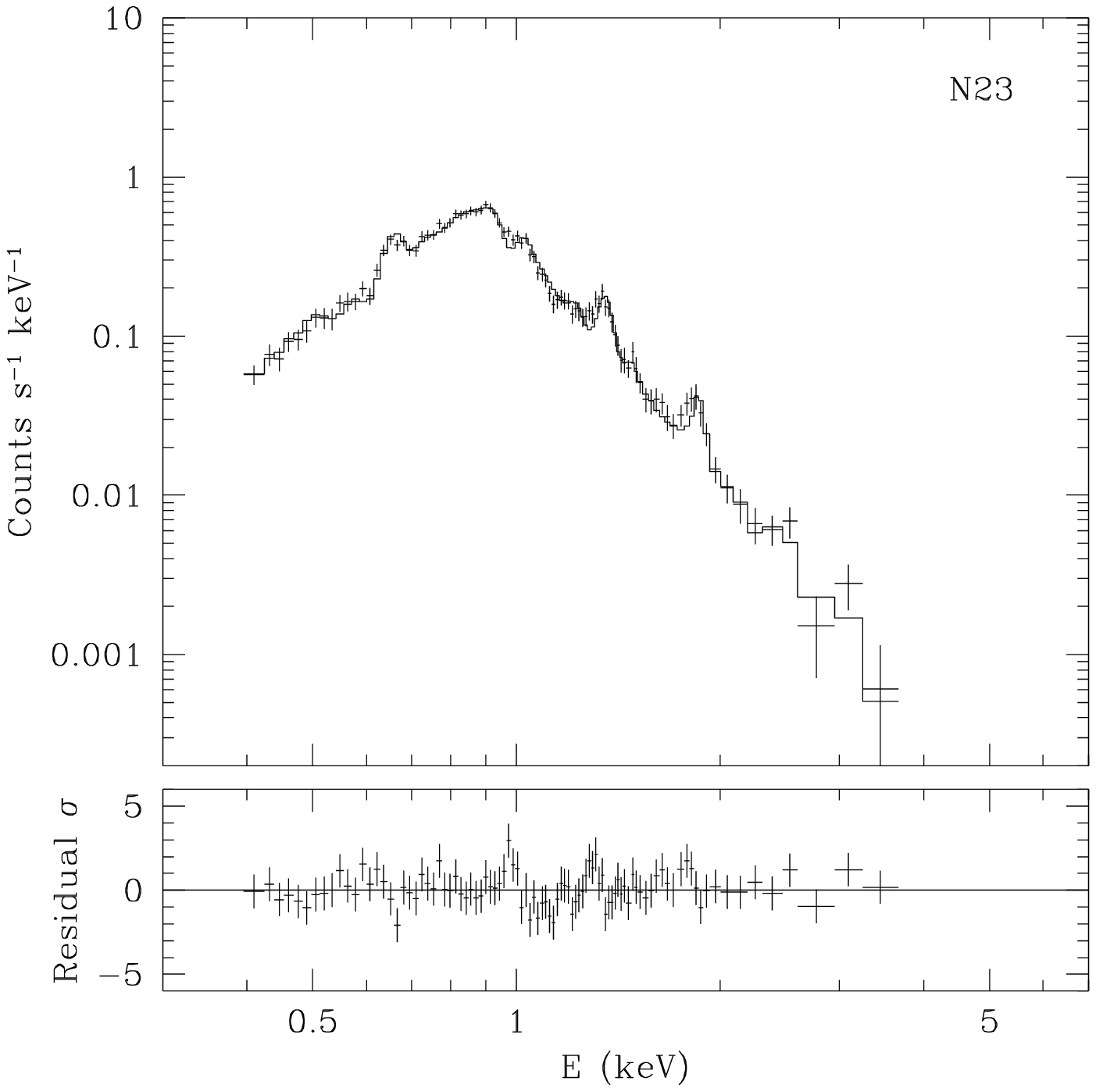}{4in}{0}{100}{100}{-300}{-200}
\end{figure}

\clearpage
\begin{figure}
\plotfiddle{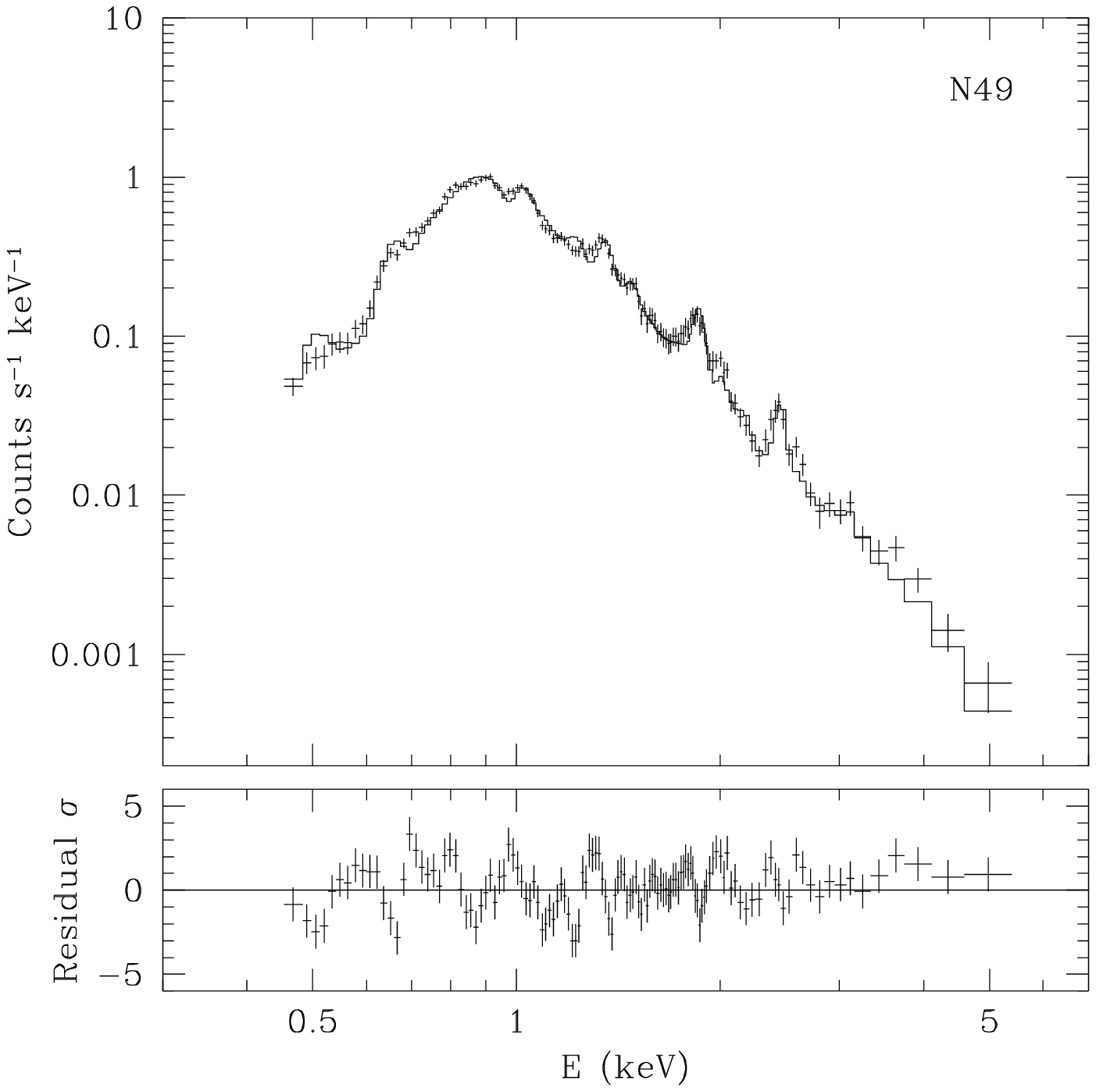}{4in}{0}{100}{100}{-300}{-200}
\end{figure}

\clearpage
\begin{figure}
\plotfiddle{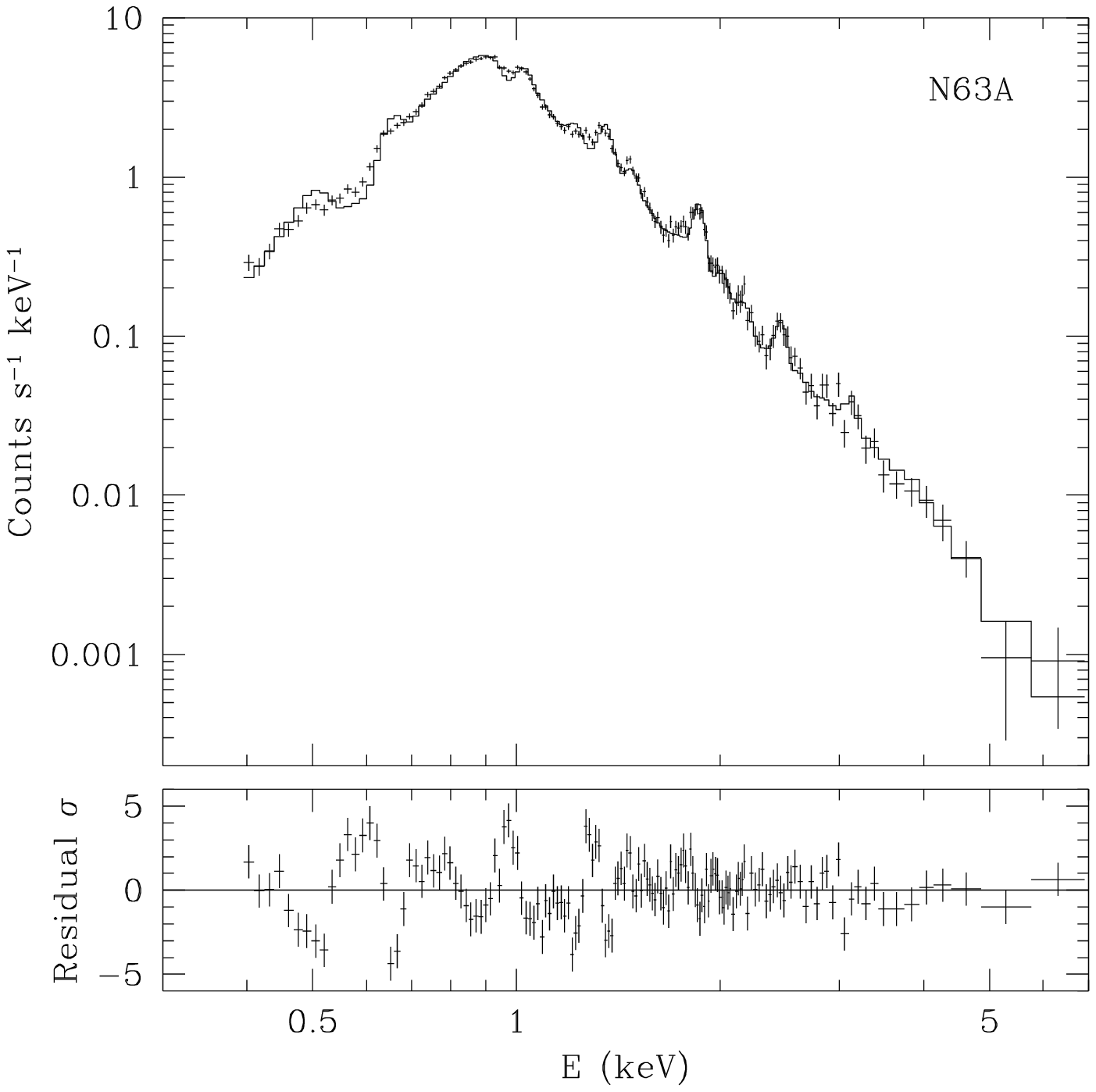}{4in}{0}{100}{100}{-300}{-200}
\end{figure}

\clearpage
\begin{figure}
\plotfiddle{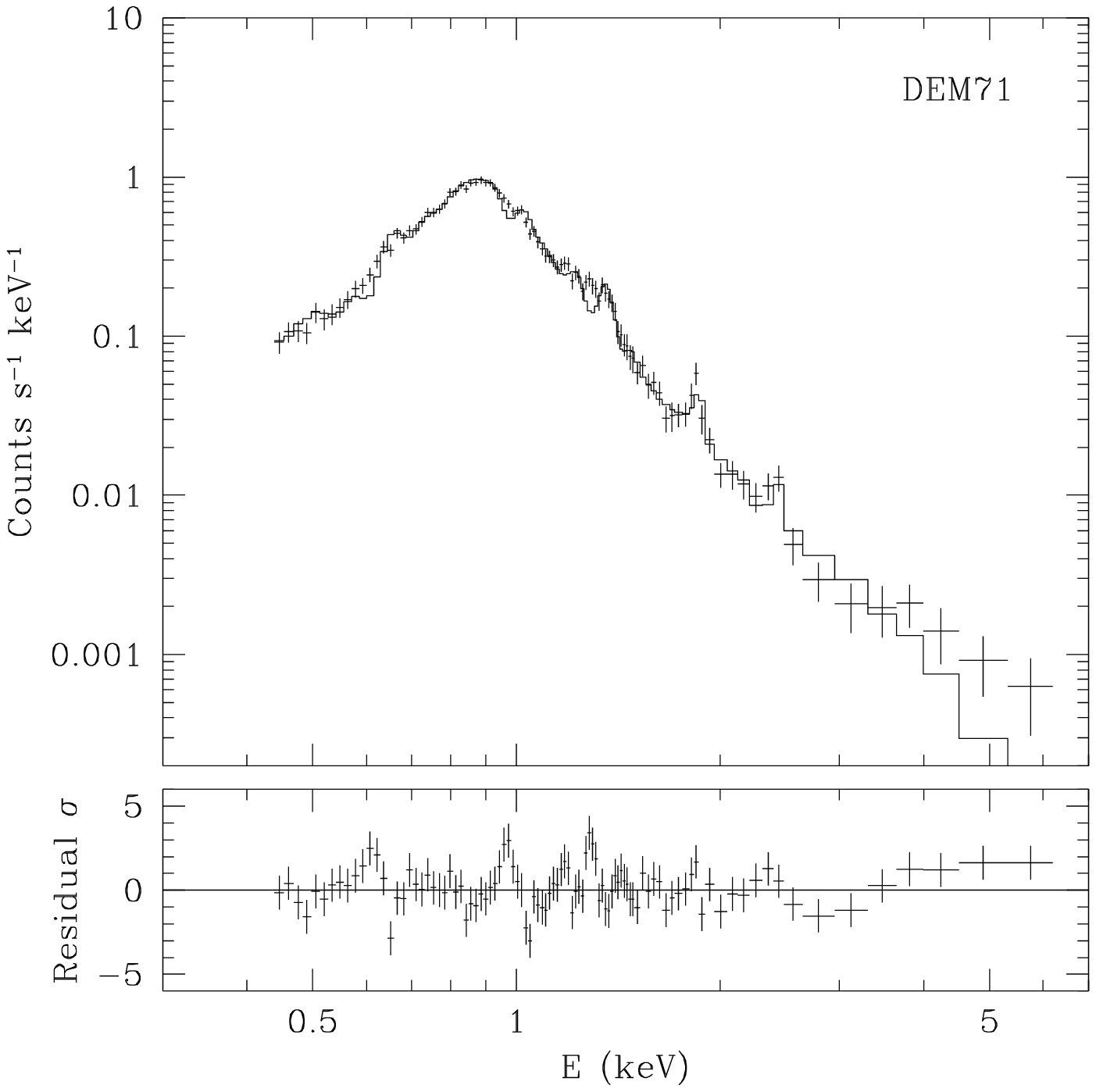}{4in}{0}{100}{100}{-300}{-200}
\end{figure}

\clearpage
\begin{figure}
\plotfiddle{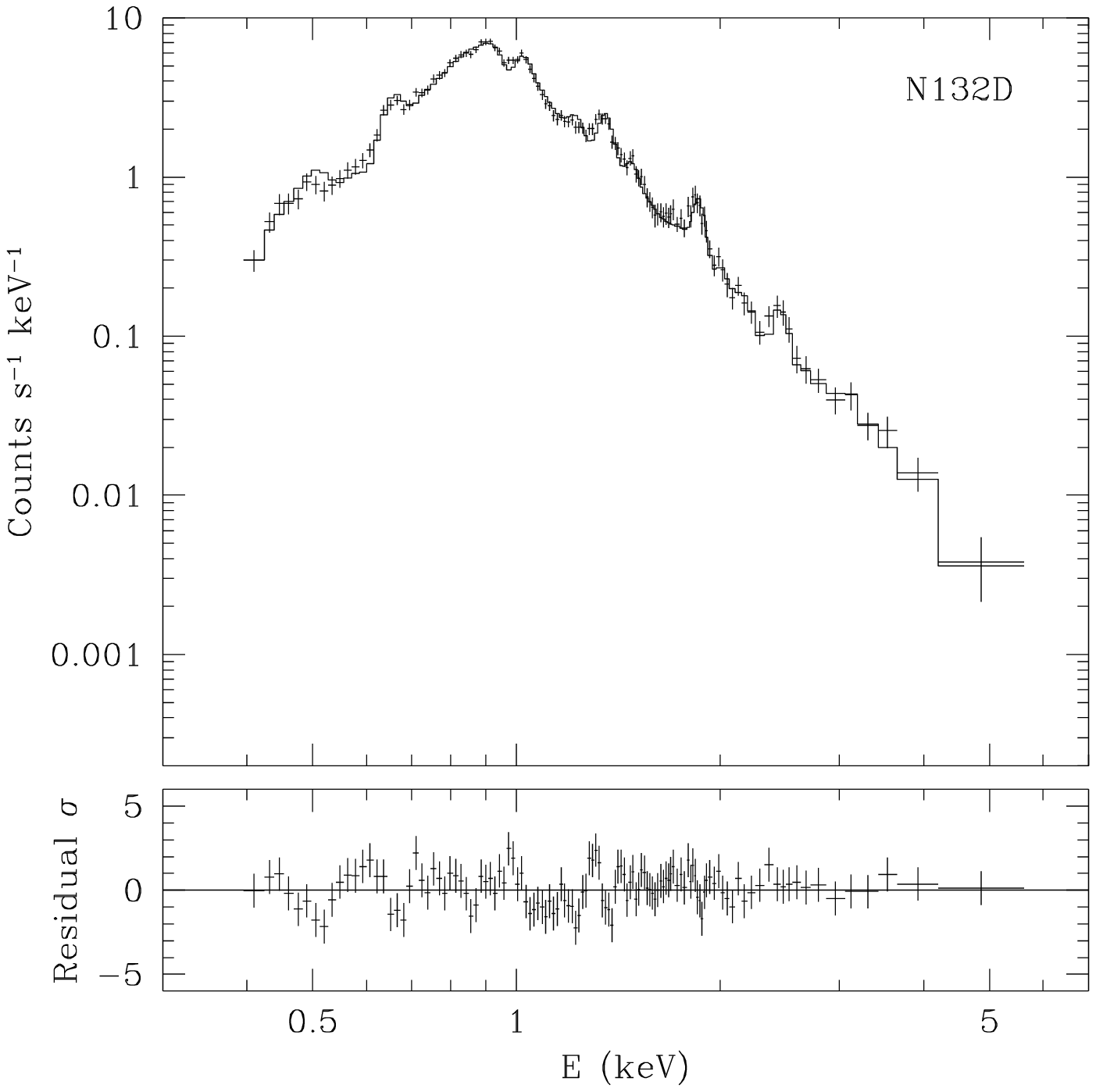}{4in}{0}{100}{100}{-300}{-200}
\end{figure}

\clearpage
\begin{figure}
\plotfiddle{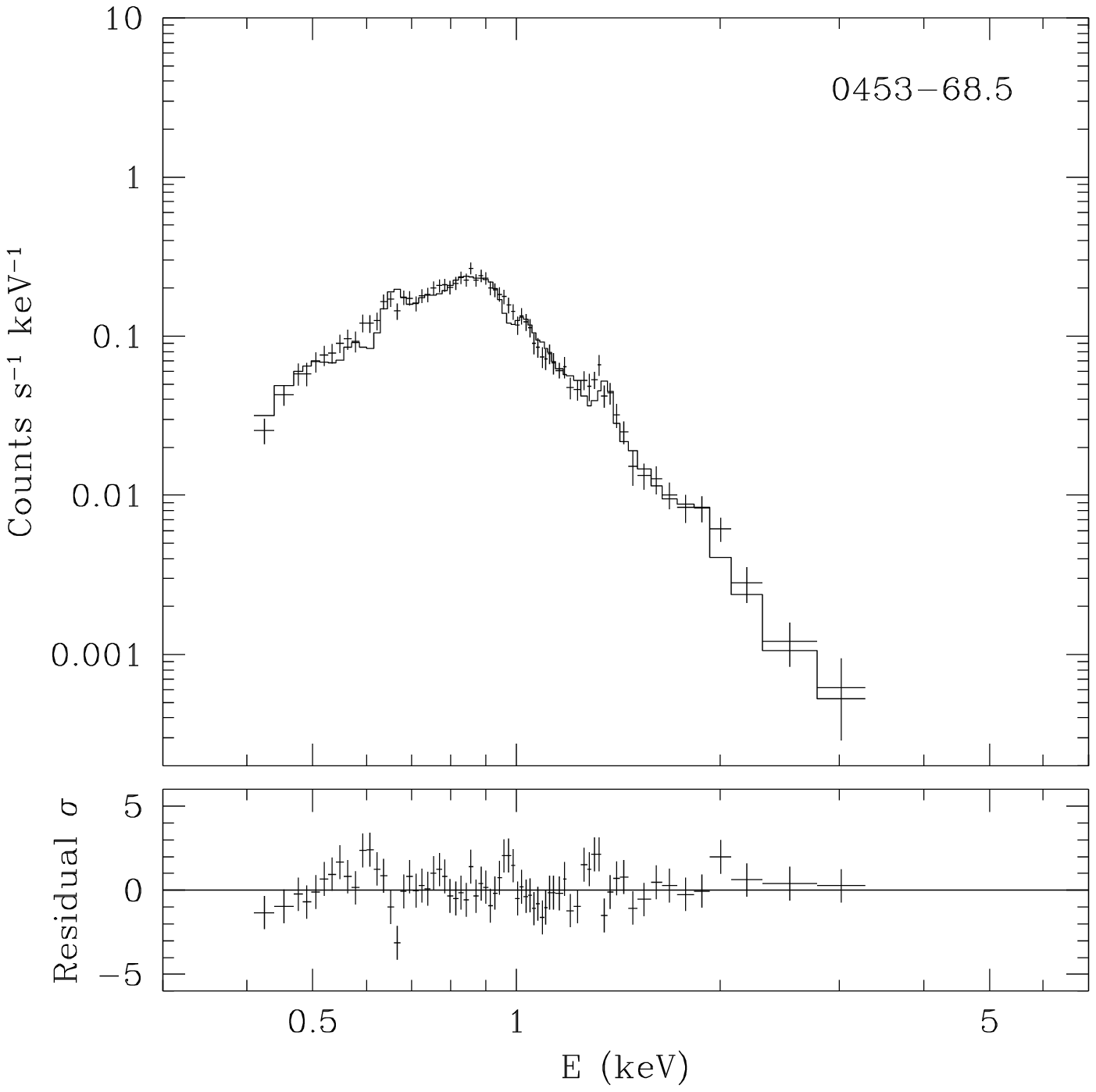}{4in}{0}{100}{100}{-300}{-200}
\end{figure}

\clearpage
\begin{figure}
\plotfiddle{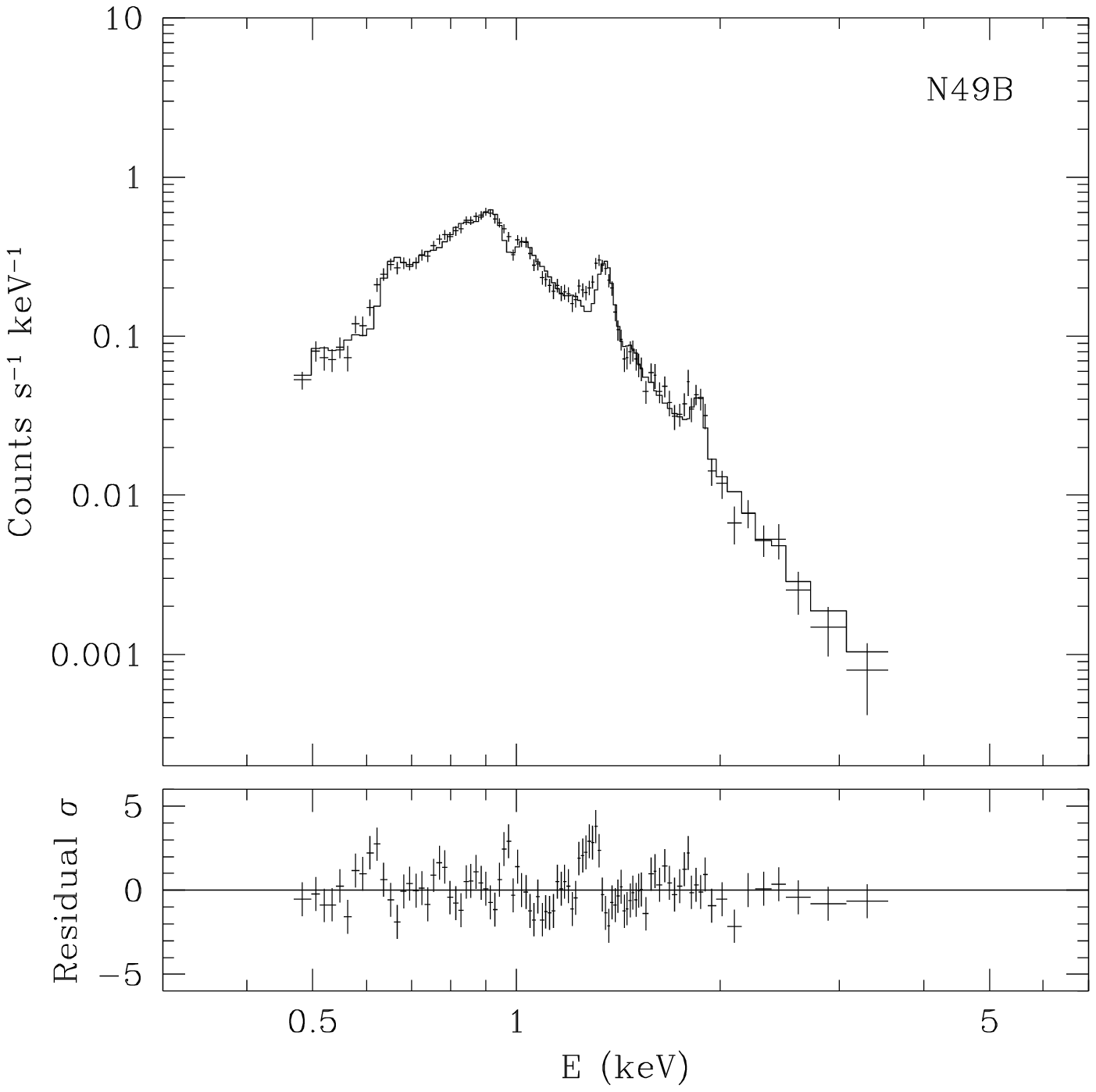}{4in}{0}{100}{100}{-300}{-200}
\end{figure}

\clearpage
\begin{figure}
\plotfiddle{fig2.ps}{4in}{0}{100}{100}{-300}{-200}
\end{figure}

\clearpage
\begin{figure}
\plotfiddle{fig3.ps}{4in}{0}{100}{100}{-300}{-250}
\end{figure}

\clearpage
\begin{figure}
\plotfiddle{fig4.ps}{4in}{0}{100}{100}{-300}{-250}
\end{figure}

\clearpage
\begin{figure}
\plotfiddle{fig5.ps}{4in}{0}{100}{100}{-300}{-250}
\end{figure}

\clearpage
\begin{figure}
\plotfiddle{fig6.ps}{4in}{0}{100}{100}{-300}{-250}
\end{figure}

\end{document}